\documentclass[twocolumn]{openjournal}

\usepackage[T1]{fontenc}
\usepackage[utf8]{inputenc}
\usepackage{xcolor}
\usepackage{hyperref}
\usepackage{orcidlink}

\definecolor{linkcolor}{rgb}{0.0,0.3,0.5}
\hypersetup{
    unicode=true,
    colorlinks=true,
    linkcolor=linkcolor,
    citecolor=linkcolor,
    filecolor=linkcolor,
    urlcolor=linkcolor
}

\makeatletter
\clo@superscriptaddress
\def\frontmatter@authorformat{\footnotesize\scshape\centering}
\def\frontmatter@affiliationfont{\normalfont\scriptsize}
\def\frontmatter@abstractfont{\small\parindent=0.125in}
\def\frontmatter@title@below{\vspace*{0.12in}}
\makeatother

\begin{document}

\title{\textit{JWST} and Keck observations of the off-nuclear tidal disruption event TDE~\lowercase{2025abcr}: An evolving reprocessing layer}

\author{Kishore C. Patra\orcidlink{0000-0002-1092-6806}}
\affiliation{Department of Astronomy and Astrophysics, UC Santa Cruz, 1156 High Street, Santa Cruz, CA 95064 }
\affiliation{University of California Observatories, 1156 High Street, Santa Cruz, CA 95064}
\email{kcpatra@ucsc.edu}

\author{Emily R. Liepold\orcidlink{0000-0002-7703-7077}}
\affiliation{Department of Astronomy, University of California, Berkeley, CA 94720, USA }

\author{Nicholas Earl\orcidlink{0000-0003-1714-7415}}
\affiliation{Department of Astronomy, University of Illinois at Urbana-Champaign, 1002 W. Green St., IL 61801, USA}

\author{Ryan J. Foley\orcidlink{0000-0002-2445-5275}}
\affiliation{Department of Astronomy and Astrophysics, UC Santa Cruz, 1156 High Street, Santa Cruz, CA 95064 }

\author{Chung-Pei Ma\orcidlink{0000-0002-4430-102X}}
\affiliation{Department of Astronomy, University of California, Berkeley, CA 94720, USA }
\affiliation{Department of Physics, University of California, Berkeley, CA 94720, USA}

\author{Sebastian Gomez\orcidlink{0000-0001-6395-6702}}
\affiliation{Center for Astrophysics \textbar{} Harvard \& Smithsonian, 60 Garden Street Cambridge, MA 02138}
\affiliation{University of Texas Austin, 2515 Speedway, Stop C1400
Austin, TX 78712, USA}

\author{Kyle~W.~Davis\orcidlink{0000-0002-5680-4660}}
\affiliation{Department of Astronomy and Astrophysics, UC Santa Cruz, 1156 High Street, Santa Cruz, CA 95064 }

\author{Enrico~Ramirez-Ruiz\orcidlink{0000-0003-2558-3102}}
\affiliation{Department of Astronomy and Astrophysics, UC Santa Cruz, 1156 High Street, Santa Cruz, CA 95064 }

\author{K. Decker French\orcidlink{0000-0002-4235-7337}}
\affiliation{Department of Astronomy, University of Illinois at Urbana-Champaign, 1002 W. Green St., IL 61801, USA}

\author{Jonelle L. Walsh\orcidlink{0000-0002-1881-5908}}
\affiliation{George P. and Cynthia Woods Mitchell Institute for Fundamental Physics and Astronomy, and Department of Physics and Astronomy, Texas A\&M University,
College Station, TX 77843, USA}

\author{Ravjit Kaur\orcidlink{0009-0005-1871-7856}}
\affiliation{Department of Astronomy and Astrophysics, UC Santa Cruz, 1156 High Street, Santa Cruz, CA 95064 }

\author{Kirsty Taggart\orcidlink{0000-0002-5748-4558}}
\affiliation{Department of Astronomy and Astrophysics, UC Santa Cruz, 1156 High Street, Santa Cruz, CA 95064 }

\author{Joshua Candanoza\orcidlink{0009-0006-3400-9347}}
\affiliation{Department of Astronomy and Astrophysics, UC Santa Cruz, 1156 High Street, Santa Cruz, CA 95064 }

\author{V.~Ashley~Villar\orcidlink{0000-0002-5814-4061}}
\affiliation{The NSF AI Institute for Artificial Intelligence and Fundamental Interactions}
\affiliation{Center for Astrophysics \textbar{} Harvard \& Smithsonian, 60 Garden Street Cambridge, MA 02138}

\author{Prasiddha Arunachalam\orcidlink{0000-0002-6688-3307}}
\affiliation{Department of Astronomy and Astrophysics, UC Santa Cruz, 1156 High Street, Santa Cruz, CA 95064 }

\author{Phillip Macias\orcidlink{0000-0002-9946-4635}}
\affiliation{Department of Astronomy and Astrophysics, UC Santa Cruz, 1156 High Street, Santa Cruz, CA 95064 }

\author{Samaporn Tinyanont\orcidlink{0000-0002-1481-4676}}
\affiliation{National Astronomical Research Institute of Thailand, 260 Moo 4, Donkaew, Maerim, Chiang Mai, 50180, Thailand}



\begin{abstract}

Off-nuclear tidal disruption events (TDEs) provide a rare probe of massive black holes (MBHs) outside galactic nuclei. Only a handful are known, including five X-ray--selected candidates and three optically selected events. We present observations of \object{TDE~2025abcr}, the second optically selected off-nuclear TDE, discovered at a projected offset of $9.08 \pm 0.02$\,kpc from the nucleus of its host galaxy. We analyze X-ray, ultraviolet, optical, and infrared data from \textit{Swift}, Keck, the Zwicky Transient Facility, and the \textit{James Webb Space Telescope}. Broad hydrogen and helium emission lines in the optical and infrared confirm a TDE-H-He classification. From luminosity scaling relations and \texttt{MOSFiT} modeling, we infer a BH mass of $10^{6}$--$10^{7}\,M_{\odot}$, substantially smaller than the $10^{8.38 \pm 0.42}\,M_{\odot}$ BH inferred for the host-galaxy nucleus. We observe velocity evolution in the \ion{N}{3}+\ion{He}{2} emission complex, shifting from $-100$\,km\,s$^{-1}$ at day $-7$ to $+1000$\,km\,s$^{-1}$ by day $+29$, which we interpret as radiative transfer effects in an evolving reprocessing layer. The infrared spectral-energy distribution deviates from a thermal blackbody, with $\nu L_{\nu} \propto \lambda^{-2.13 \pm 0.04}$, significantly shallower than the Rayleigh--Jeans slope of $\lambda^{-3}$. We rule out dust as the source of this infrared excess. Two possibilities remain: free--free emission from reprocessing gas, or an unresolved stellar cluster at the TDE location. Reprocessed emission provides a natural explanation for the infrared excess but an underlying stellar cluster of mass $\log(M_{*}/M_{\odot}) = 7.57 \pm 0.02$ and age $<$2\,Gyr is also consistent with the data. If interpreted as a stellar cluster, the inferred mass suggests a stripped remnant of a satellite galaxy. 
The wandering MBH most likely originated in a minor merger with a smaller galaxy, although dynamical ejection from the host nucleus cannot yet be ruled out.

\end{abstract}



\maketitle 

\section{Introduction}
\label{sec:intro}

Hierarchical galaxy formation, in which structure builds up through successive mergers of smaller systems, is a fundamental prediction of cold dark matter cosmology \citep[e.g.,][]{1984Natur.311..517B, 1991ApJ...379...52W}. 
As galaxies merge, their central black holes do not always immediately settle into the nucleus of the remnant system, leaving behind a population of offset or ``wandering'' massive black holes (MBHs) within galaxy halos   \citep[e.g.,][]{Tremmel_2018_wandMBH, Ricarte_2021_wandBHB}. 
Although most of these wanderers are effectively invisible, they can occasionally reveal themselves through transient accretion events, particularly tidal disruption events (TDEs), when a star passes too close to the MBH and is tidally disrupted \citep[e.g.,][]{Evans_Kochanek_1989, Magorrian_1999_TDE-NSC, Ulmer_1999, RamirezRuiz_Rosswog_2009, MacLeod_etal_2016}.

TDEs occur when a star approaches an MBH within its tidal radius, \(R_T \approx R_\star (M_\bullet/M_\star)^{1/3}\), such that the pericenter distance \(R_p\) satisfies \(R_p \lesssim R_T\), or equivalently the penetration (impact) parameter \(\beta \equiv R_T/R_p \gtrsim 1\) \citep{Hills_1975, Rees_1988}. Roughly half of the stellar debris remains bound and eventually accretes, producing a luminous flare observable across the electromagnetic spectrum \citep{Guillochon_Ramirez_2013, Gezari_2021_tderev}.
Over the past decade, time-domain optical surveys such as the Zwicky Transient Facility \citep[ZTF;][]{Bellm_etal_2019} and All-Sky Automated Survey for SuperNovae \citep[ASAS-SN;][]{2014AAS...22323603S, 2017PASP..129j4502K} have discovered $\sim$150 TDEs to the point where population-level studies are now possible \citep{vanvelzen21_TDE_tdes, Hammerstein_2023_ZTFTDEs, Yao_Yuhan_2023_TDEs}. These events provide one of the few direct ways to study otherwise quiescent MBHs, particularly for $M_\bullet \lesssim 10^8\,M_\odot$ \citep{MacLeod_2012, Stone_2016_TDE_MBH, 2019_Mockler}.

Nearly all optically discovered TDEs have been associated with the nuclei of their host galaxies, reflecting both the central presence of MBHs \citep{Kormendy_OffsetMBHreview_2013} and the selection biases inherent in existing TDE search strategies \citep{2026arXiv260210180S}. However, in the presence of a population of wandering MBHs, a subset of TDEs is expected to occur at measurable offsets from galaxy centers. Such off-nuclear TDEs provide direct observational evidence for wandering MBHs and offer a way to probe their demographics and dynamical histories \citep{patra2025jwstkeckobservationsoffnuclear, Yao_Yuhan_2025_24tvd}. In addition, their spatial separation from the host nucleus reduces contamination from the surrounding stellar population and dust, enabling cleaner constraints on TDE accretion physics \citep{patra2025jwstkeckobservationsoffnuclear, Guolo_2026_24tvd}.

A small but growing sample of off-nuclear TDEs has now emerged. Most early candidates were identified in X-ray surveys and are often associated with faint stellar systems such as globular clusters, ultra-compact dwarf galaxies, or satellite galaxies \citep[WINGS J134849.88+263557.5, 3XMM J2150, EP240222a;][]{2013MNRAS.435.1904M, Lin_2018_J215, Lin_2020_J215, Jin_2025_EP24}.
More recently, TDE~2024tvd became the first optically selected off-nuclear TDE, with high-resolution \textit{Hubble Space Telescope} (\textit{HST}) and \textit{James Webb Space Telescope} (\textit{JWST}) imaging revealing a projected offset of 0.8~kpc from its host nucleus \citep{patra2025jwstkeckobservationsoffnuclear, Yao_Yuhan_2025_24tvd}. Another optically selected TDE---2023mfm, initially thought to be nuclear TDE---was confirmed as an off-nuclear TDE with a 1.1\,kpc offset using late-time radio observations\citep{2026RNAAS..10..153L, Sfaradi_23mfm}.

Detailed follow-up observations of TDE~2024tvd indicate that the TDE originates from a wandering $\sim$10$^{6}\,M_{\odot}$ MBH \citep{patra2025jwstkeckobservationsoffnuclear, Yao_Yuhan_2025_24tvd, Guolo_2026_24tvd}. Of the three major possible origins for the wandering MBH, ejection from the nucleus due to gravitational-wave recoil can be ruled out, although dynamical ejection via three-body interactions remains a possibility \citep{Volonteri_Perna_2005_ejectMBH, Komossa_2012_recoilBHrev, 2016MNRAS.456..961B}. The most likely origin is a past minor galaxy merger \citep[e.g.][]{Tremmel_2018_wandMBH}, supported by the lack of tidal tails or kinematic signatures of a major merger, as well as the possible presence of a compact stellar system, such as a nuclear star cluster (NSC) or a stripped satellite galaxy, around the MBH (\citealt{patra2025jwstkeckobservationsoffnuclear}, although see \citealt{Guolo_2026_24tvd}).

\begin{figure}[tbp]
    \centering
    \includegraphics[width=\linewidth]{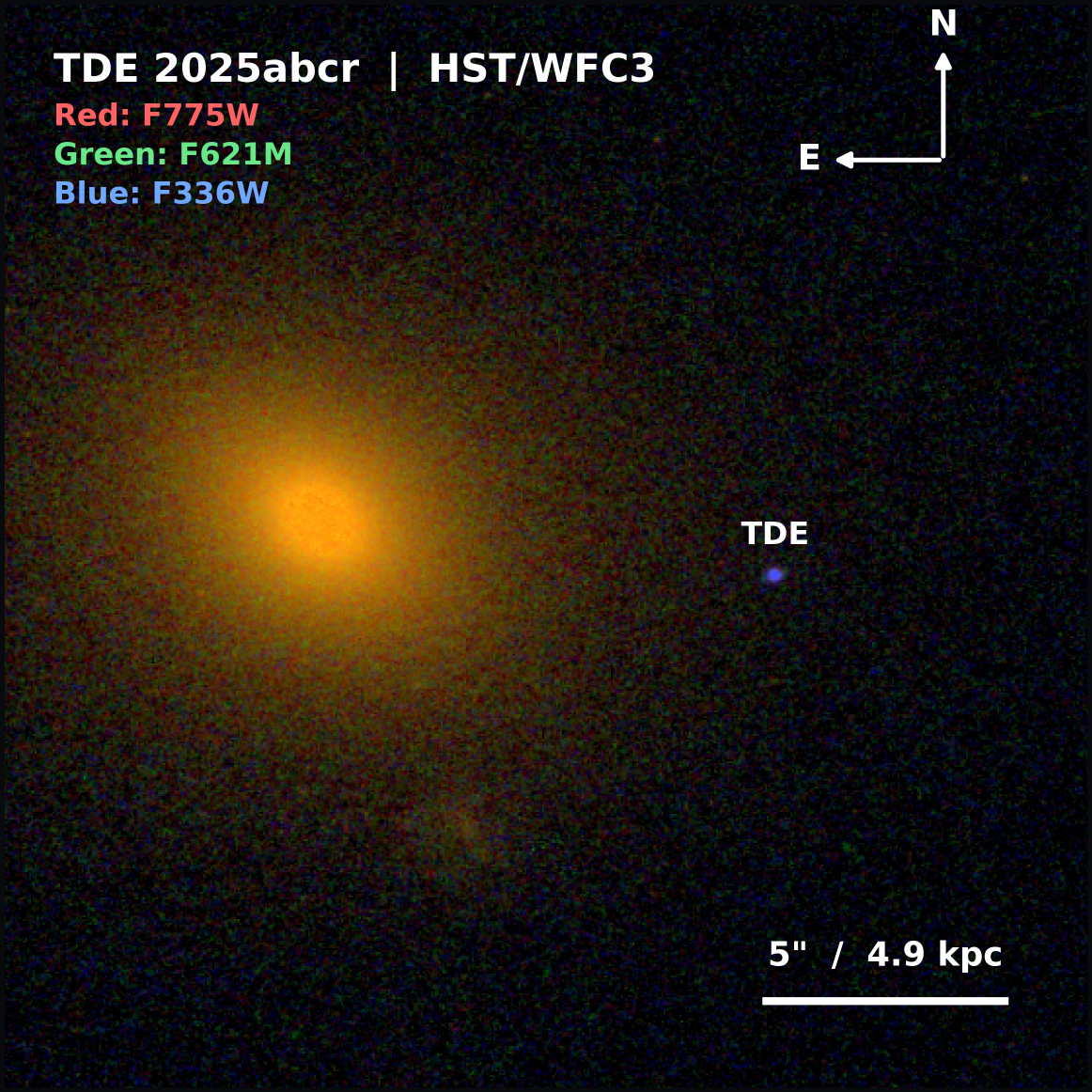}
    \caption{High spatial resolution image of TDE~2025abcr and its host galaxy developed by combining publicly available \textit{HST} images. The TDE stands out as a blue point source in the outskirts of a red early-type galaxy.}
    \label{fig:hstimg}
\end{figure}

In this work, we present \textit{JWST} and Keck observations of the second optically selected off-nuclear TDE~2025abcr (Figure \ref{fig:hstimg}). It was discovered in the ZTF alert stream and spectroscopically classified by \citet{Stein_2025abcr_2025TNS} as an off-nuclear TDE on UT 2025-Nov-09. The transient was selected by a systematic search for off-nuclear TDEs with an implementation of the machine-learning classifier \texttt{tdescore} that was agnostic to the location of the TDE within the galaxy \citep{2024ApJ...965L..14S,2026arXiv260210180S}.
The independent multi-wavelength analysis of \citet{2026arXiv260210180S} likewise established a TDE interpretation despite the event's unusual, highly offset location.

The key questions raised by TDE~2025abcr are similar to those posed by earlier off-nuclear events: what is the origin of the offset MBH, and what stellar reservoir enables TDE production at such large galactocentric distances? The former could reflect processes such as gravitational recoil, dynamical ejection, or galaxy mergers, while the latter likely depends on the presence of a dense stellar system such as an NSC, which can enhance the TDE rate by increasing the local stellar phase-space density and facilitating efficient loss-cone refilling \citep{Magorrian_1999_TDE-NSC, Stone_2016_TDE_MBH, Pfister_2020_tderate_highz, Melchor_2024}. To address these questions, we combine multi-wavelength observations spanning X-ray, ultraviolet (UV), optical, and infrared (IR) bands 
and conduct spectral energy distribution (SED) modeling of the TDE.

The paper is organized as follows: in Section \ref{sec:obs} we describe the observations of TDE~2025abcr. Section \ref{sec:analysis_results} presents the data analysis and modeling of the TDE and host galaxy. In Section \ref{sec:discussion} we discuss the results and their implications. Finally, in Section \ref{sec:conclusion} we summarize our conclusions. We use UT time throughout this paper and assume a flat $\Lambda$CDM cosmology with $H_{0} = 70~\mathrm{km~s^{-1}~Mpc^{-1}}$, $\Omega_{\rm m} = 0.3$, and $\Omega_{\Lambda} = 0.7$.

\section{Observations}
\label{sec:obs}

\subsection{X-ray}

    \begin{figure*}[tbp]
        \centering
    	\includegraphics[width=0.7\linewidth]{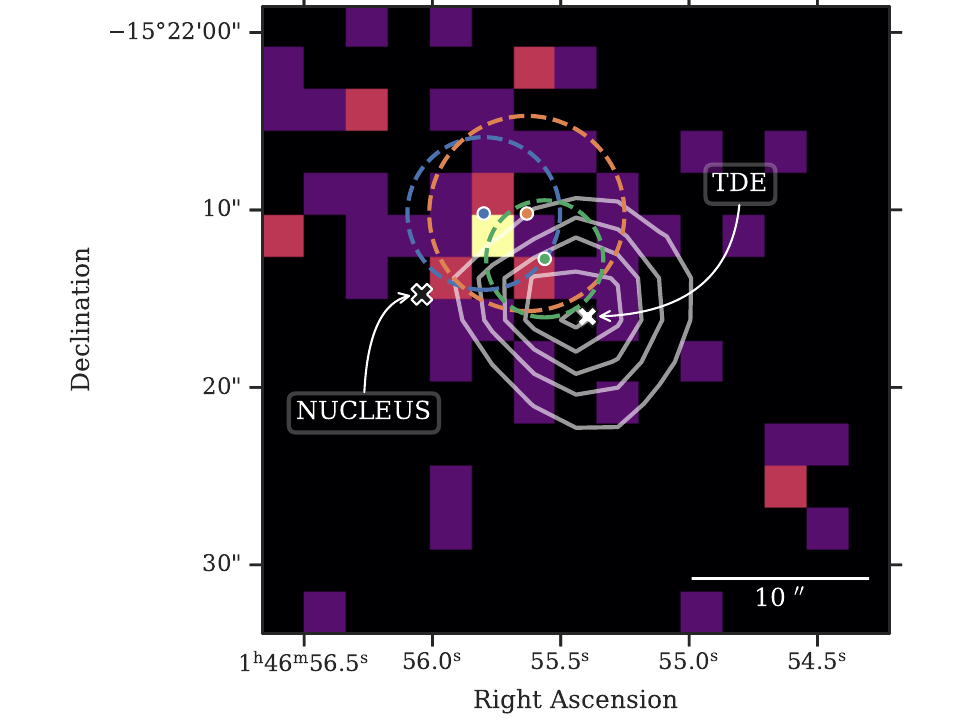}
    	\caption{ Stacked \textit{Swift} XRT image of the region around TDE~2025abcr. The positions of the host galaxy nucleus (black) and the TDE (white) are marked with ``X''. White contours represent the \textit{Swift} UVOT flux, which is consistent with the TDE position. The astrometric position (colored dots) and uncertainty (dashed circles) of the X-ray source derived using three methods are shown: the standard XRT position based on \textit{Swift} star-tracker astrometry (blue), the UVOT-enhanced astrometric solution (green), and the position obtained by aligning XRT sources with the 2MASS catalog (orange). Given the current data, the origin of the X-ray emission is unclear; it may originate from the TDE, the host nucleus, or another source within the galaxy.}
        \label{fig:Xray_source}
    \end{figure*}

    \begin{figure*}[tbp]
        \centering
    	\includegraphics[width=\linewidth]{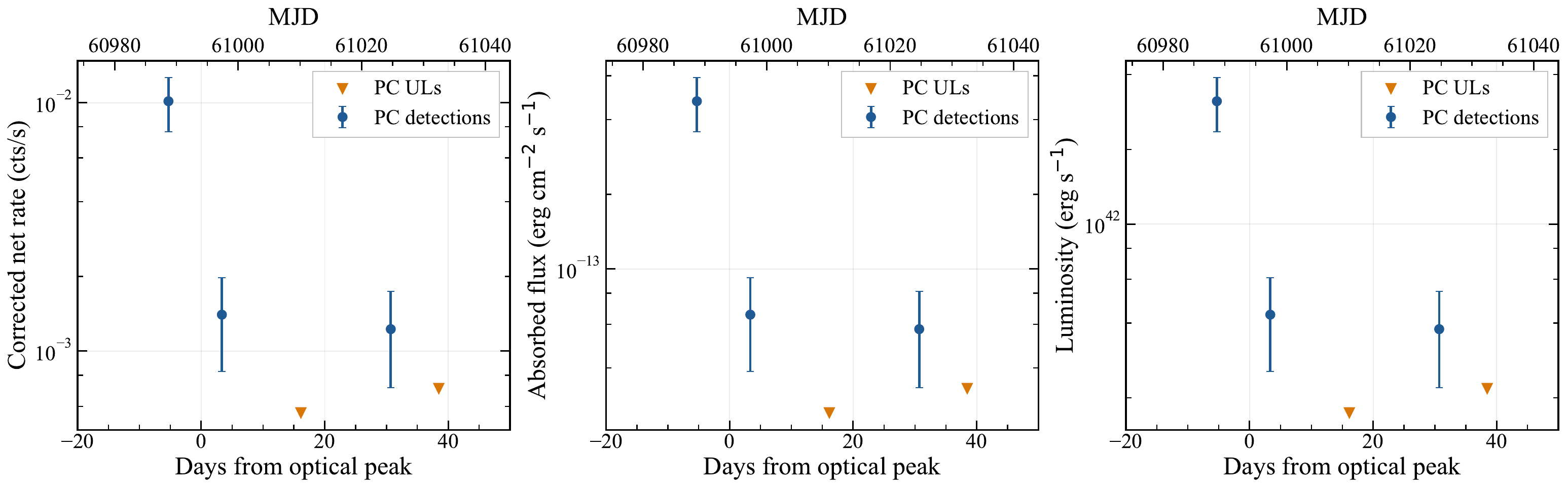}
    	\caption{X-ray light curve observed near the TDE location. The X-ray flux rapidly declines and becomes undetectable after $+40$ days.}
        \label{fig:xrtlc}
    \end{figure*}

X-ray observations were obtained with the \textit{Neil Gehrels Swift Observatory} X-ray Telescope \citep[XRT;][]{Burrows_etal_2005}. The data were reduced using the \texttt{HEASoft} software package (v6.35) and the latest calibration files from \texttt{CALDB}. Event files were reprocessed with \texttt{xrtpipeline}, and source products were generated using \texttt{xrtproducts}.

Source counts were extracted using a circular aperture of radius $50''$ centered on the X-ray source identified in the XRT image. Background counts were extracted from a nearby source-free region using an annular aperture to mitigate contamination from diffuse emission and field sources. Exposure maps and ancillary response files were generated with standard tools, and the appropriate response matrix files were applied.

To improve signal-to-noise ratio (SNR), we constructed a stacked spectrum by combining all available observations. The resulting spectrum was grouped to ensure a minimum number of counts per bin prior to fitting. Spectral modeling was performed in \texttt{XSPEC} \citep{Arnaud_1996_XSPEC}.

The spectrum was fitted with an absorbed, redshifted Comptonized disk model, \texttt{TBabs*zashift*simpl*ezdiskbb}. The Galactic absorption was fixed at $N_{\rm H}=1.55\times10^{20}\ {\rm cm^{-2}}$ \citep{2016_HIPI4}, the redshift was fixed at $z=0.04985$ \citep{Jones_2009_6dF}, and the photon index was fixed at $\Gamma=2.5$ \citep{Mummery_Balbus_2021}. The fit yielded $\chi^2 \approx 107$ for 89 degrees of freedom. No additional intrinsic absorption component was required by the fit, and we find no evidence for neutral hydrogen absorption beyond the Galactic column. The best-fit model yields a scattering fraction $f_{\rm sc}=0.41_{-0.23}^{+0.28}$ and disk temperature $kT_{\max}=0.057_{-0.024}^{+0.047}$~keV. The corresponding absorbed 0.3--10~keV flux is $F_X = 6.93_{-0.92}^{+0.99}\times10^{-14}~\rm erg~cm^{-2}~s^{-1}$, corresponding to an absorbed luminosity of $L_X = 4.16_{-0.57}^{+0.61}\times10^{41}~\rm erg~s^{-1}$. The large uncertainties on the disk normalization reflect the limited photon statistics of the stacked spectrum.

Count rates were binned to a minimum number of counts per bin and corrected for exposure and background. These corrected count rates were converted to absorbed 0.3--10~keV fluxes using a single counts-to-flux conversion factor derived from the best-fit model of the stacked spectrum, and corresponding luminosities were computed assuming a luminosity distance given by $z=0.04985$.

We detect a faint X-ray source offset by $6.7''$ from the TDE. However, given the typical centroiding uncertainty of XRT ($\sim$2-3$''$), it is not possible to determine whether the emission originates from the TDE, the host galaxy nucleus, or another source within the galaxy. A stacked image combining all X-ray observations between 2025-Nov-09 and 2026-Jan-16 reveals an X-ray source whose localization, within astrometric uncertainties, is offset from both the TDE and the host nucleus. Figure~\ref{fig:Xray_source} shows the X-ray source localization relative to the TDE and the host galaxy. We derived the X-ray source position using three astrometric methods: the standard XRT solution based on \textit{Swift} star-tracker astrometry, the UVOT-enhanced solution, and a solution obtained by aligning XRT sources to the 2MASS reference frame. All three methods showed an offset between the TDE location and the X-ray source.
Owing to this ambiguity, we do not include the X-ray data in the SED analysis (Section \ref{sec:sed_modeling}).

Nonetheless, the measured X-ray luminosity in the 0.3-10 keV band ranges between $(7-25)\times10^{41} ~\rm erg~s^{-1}$  and shows a decline over time, eventually falling below the XRT detection threshold around day $+40$ (Figure~\ref{fig:xrtlc}). We combined all available observations to construct a stacked X-ray spectrum (shown in Figure~\ref{fig:sed}). The spectrum exhibits both a soft and a hard component. The soft emission may be associated with the TDE accretion flow, while the hard component could come from low-level activity of the host galaxy nucleus \citep[e.g.,][]{Auchettl_2017}. However, without higher spatial resolution observations (for example with \textit{Chandra}), this interpretation remains uncertain. This soft-plus-hard spectral structure, the rapid fading of the soft component, and the inferred X-ray luminosity are broadly consistent with the independent analysis of \citet{2026arXiv260210180S}. However, they attribute the soft X-ray component to the TDE and note that the harder component may include unresolved host-galaxy emission.

\subsection{UV}

    \begin{figure*}[tbp]
        \centering
    	\includegraphics[width=0.6\linewidth]{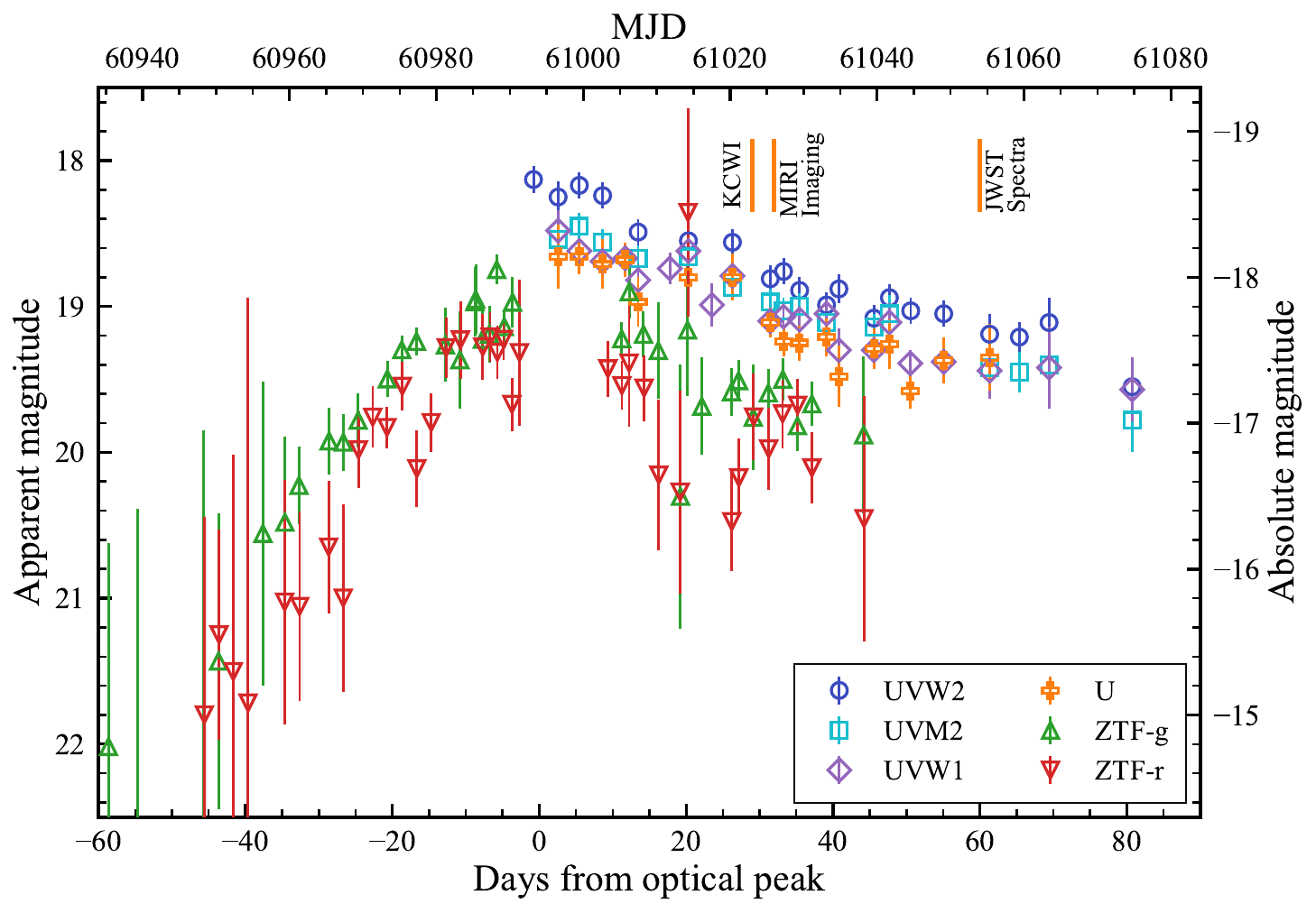}
    	\caption{UV and optical light curves of TDE~2025abcr. The epochs of KCWI and \textit{JWST} observations are marked.}
       \label{fig:uvopt_lc}
    \end{figure*}

UV observations of TDE~2025abcr were obtained with the \textit{Swift} Ultraviolet/Optical Telescope \citep[UVOT;][]{Roming_etal_2005} in the $UVW2$, $UVM2$, $UVW1$, and $U$ bands (Figure \ref{fig:uvopt_lc}). The data were reduced using the standard \texttt{uvotsource} task, with a $5''$ source aperture and a larger nearby background region. The UVOT fluxes were corrected for Galactic extinction assuming $E(B-V) = 0.013$~mag \citep{Schlafly_etal_2011} and the \citet{Cardelli_etal_1989} extinction law with $R_{\rm V} = 3.1$. Due to the absence of pre-transient \textit{Swift} templates, we did not explicitly subtract the host-galaxy contribution. Given the large \(9.3''\) separation between the TDE and the galaxy nucleus, as well as the quiescent, UV-faint nature of the host, we estimate from the host-galaxy SED that the residual host contribution is \(\lesssim 2\%\) in all 3 UV bands. Because this contribution is substantially smaller than the photometric uncertainties of \(12\)--\(15\%\), we apply no additional correction to the UV fluxes.

\subsection{Optical}

    \begin{figure*}[tbp]
        \centering
    	\includegraphics[width=\linewidth]{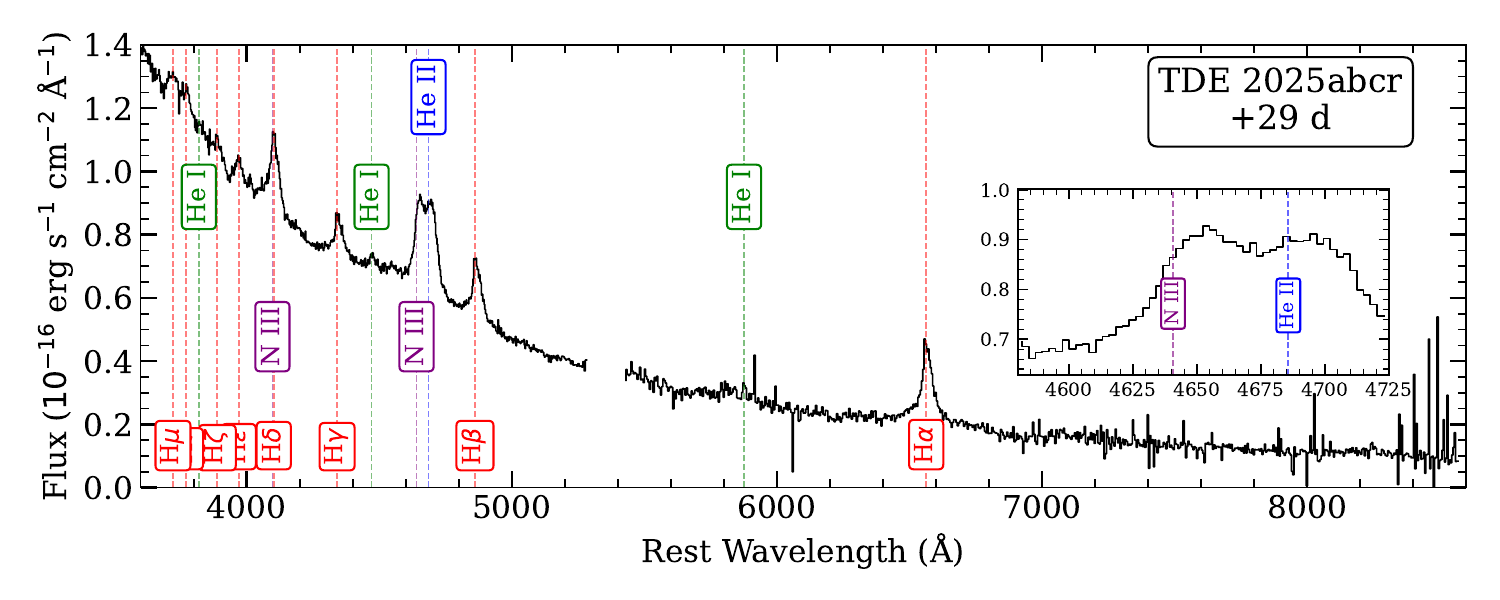}
    	\caption{Optical spectrum of TDE~2025abcr at phase +29 days observed with KCWI. For clarity, the blue and red sides have been binned by factors of 6 and 10, respectively, relative to the native $0.5$~\AA\  sampling. 
        Prominent H and He emission lines are marked. The \ion{N}{3} Bowen complex is also identified. The inset shows a zoomed-in view of the \ion{N}{3}+\ion{He}{2} lines which are redshifted by $\sim 1000$~km~s$^{-1}$. The blue continuum and spectral features are consistent with a TDE-H-He classification.
        }
       \label{fig:optical_spec}
    \end{figure*}

Publicly available ZTF $g$- and $r$-band photometry between 2025-Oct-02 and 2025-Dec-22 was used for producing light curves (Figure \ref{fig:uvopt_lc}) and in Modular Open-Source Fitter for Transients \citep[\texttt{MOSFiT};][see the Appendix]{Guillochon_2018_mosfit} analysis.

We obtained an optical spectrum of TDE~2025abcr (Figure~\ref{fig:optical_spec}) on 2025 December 13 (day $+29$) with the Keck Cosmic Web Imager \citep[KCWI;][]{Morrissey_2018_KCWI} mounted on the Keck II telescope atop Maunakea, Hawai‘i. The observations were carried out under program U208 (PI: C.-P. Ma) using both the blue and red channels, providing usable wavelength coverage of 3500--8500~\AA. The instrument was configured with the small slicer and the low-resolution BL and RL gratings, yielding average spectral resolutions of $R \sim 3600$ and $R \sim 2000$ in the blue and red channels, respectively. The field of view was $8.4'' \times 20.4''$ with a spatial sampling of $0.35''$ per spaxel.

The observations were obtained under excellent conditions, with an average airmass of 1.32 and a seeing of $0.5''$. The total exposure time was 1035\,s on the blue side, while the red channel was observed in three exposures of 285\,s each to enable effective cosmic-ray rejection. The integral field unit (IFU) data cubes were produced using the automated \texttt{KCWI DRP} pipeline \citep{Neill_2023_KCWIDRP}. Since the \texttt{KCWI DRP} pipeline produces wavelengths in vacuum by default, we converted them to air wavelengths using the prescription of \citet{Morton_1991_vac_air}.

The local sky and host-galaxy background at the TDE position was modeled with a two-dimensional polynomial and subtracted prior to flux extraction. The TDE flux was measured using an aperture with a radius of $3\times$ the full width at half maximum (FWHM) of the seeing disk. Assuming a \citet{Moffat_1969_psf} PSF under seeing-limited conditions, such an aperture encloses $\sim 98\%$ of the total point-source flux. We applied a wavelength-independent aperture correction of 2\%\footnote{Under Kolmogorov turbulence, the seeing FWHM varies weakly with wavelength as $\lambda^{-0.2}$ \citep{Fried_1966_Kolmogorov}.} and added a 2\% systematic uncertainty in quadrature to the statistical flux errors. Flux calibration was performed using a custom pipeline\footnote{\url{https://github.com/UCSC-Transients/UCSC_KCWI_spectral_pipeline}} for KCWI IFU cubes with the spectrophotometric standard star Feige\,110, which was observed with the same KCWI instrument setup \citep{patra2025jwstkeckobservationsoffnuclear}.

\subsection{Infrared}

    \begin{figure*}[tbp]
        \centering
    	\includegraphics[width=\linewidth]{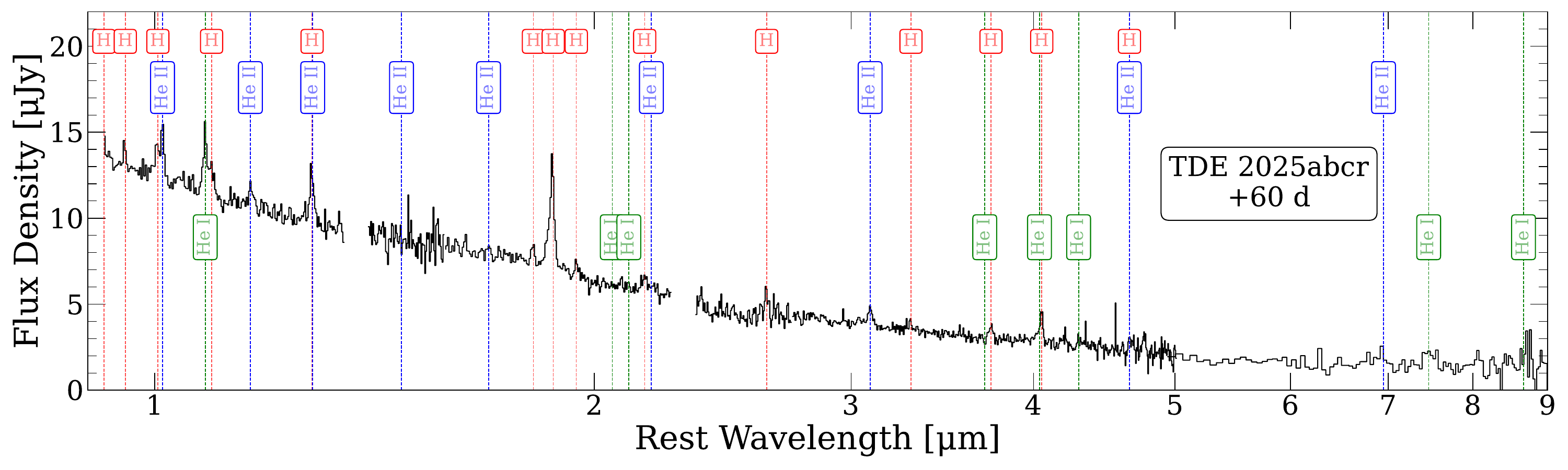}
    	\caption{IR spectrum of TDE~2025abcr at phase +60 days observed with \textit{JWST} NIRSpec and MIRI LRS. For clarity, the displayed NIRSpec spectrum is binned by a factor of 10 for the G140H/F100LP and G235H/F170LP segments and by a factor of 3 for the G395M/F290LP segment, while the MIRI LRS spectrum is shown at its native sampling. Prominent H and He emission lines are marked.}
        \label{fig:jwst_spec}
    \end{figure*}

TDE~2025abcr was observed with \textit{JWST} using the Mid-Infrared Instrument imager \citep[MIRI;][]{MIRI_2015}, the MIRI Low Resolution Spectrometer \citep[LRS;][]{MIRI_2015}, and the NIRSpec integral field unit (IFU) spectrograph \citep{NIRSpec_2022} as part of a Cycle 4 Director’s Discretionary (DD) program (ID 12485; PI: K. C. Patra). MIRI imaging (Figure \ref{fig:jwstmiri}) was obtained on 2025-Dec-16 (day $+31$) and was used to determine a precise astrometric position of the TDE. Spectroscopic observations with NIRSpec IFU and MIRI LRS (Figure \ref{fig:jwst_spec}) were carried out on 2026 January 14 (day $+60$). 

\subsubsection{MIRI imaging}

MIRI imaging was performed in the F560W filter using a large cycling dither pattern with 4 dither positions. The observations used the FASTR1 readout with 20 groups per integration, 3 integrations per exposure, and one exposure per dither position, resulting in a total of 12 integrations and a total exposure time of 688\,s. The MIRI data were reduced using the standard \textit{JWST} Calibration Pipeline \citep[version 13.0.6;][]{Bushouse_2023_pipeline}, with an appropriate CRDS context (\texttt{jwst\_1464.pmap}). Stage~1 processing applied detector-level corrections, including reference pixel subtraction, non-linearity correction, dark current subtraction, and ramp fitting to produce count-rate images. Stage~2 applied flat-fielding, flux calibration, and WCS assignment, resulting in calibrated (\texttt{\_cal}) products. Stage~3 combined the dithered exposures using the drizzle algorithm to produce a final distortion-corrected mosaic (\texttt{\_i2d}).

\subsubsection{NIRSpec IFU}

NIRSpec IFU observations were obtained using a 4-point dither pattern to improve spatial sampling and mitigate detector systematics. Three grating/filter combinations were used: G140H/F100LP, G235H/F170LP, and G395M/F290LP, providing spectral coverage from $\sim$1--5~$\mu$m at medium-to-high spectral resolution. The observations used the NRSIRS2 readout pattern with 12, 18, and 15 groups per integration for the G140H, G235H, and G395M settings, respectively, with one integration per exposure at each dither position. This resulted in total exposure times of 3560\,s (G140H), 5310\,s (G235H), and 4435\,s (G395M).

Following \citet{patra2025jwstkeckobservationsoffnuclear}, the data were reduced using the standard \textit{JWST} Calibration Pipeline steps. The TDE spectrum was extracted using a custom aperture photometry procedure applied to the final data cubes \citep[see][for details]{patra2025jwstkeckobservationsoffnuclear}. For each wavelength slice, a $1.2'' \times 1.2''$ cutout centered on the TDE was constructed. The host galaxy contribution within this region was modeled as a smooth two-dimensional polynomial surface, masking a central circular region of radius $0.3''$ to exclude the TDE emission. 
This model was subtracted to isolate the TDE flux, which was then measured within a circular aperture of radius $0.2''$. A wavelength-dependent aperture correction was applied based on synthetic point spread function (PSF) cubes generated with the \texttt{STPSF} package \citep{Perrin_2025_STPSF}. The resulting spectrum was converted from vacuum to air wavelengths following \citet{Morton_1991_vac_air} and shifted to the host-galaxy rest frame at $z = 0.04985$.

\subsubsection{MIRI LRS}

The TDE coordinates were first determined to $<0.1''$ accuracy using the previously obtained high-spatial-resolution MIRI image. MIRI LRS observations were then acquired using a bright offset star at (RA, Dec) = (01:46:54.2745, $-15$:22:14.50), which was centered in the LRS slit before applying the offset to the TDE position. The acquisition was verified with a pointing verification exposure in the F560W filter prior to the science exposures. The LRS observations were performed using the FASTR1 readout pattern with 35 groups per integration and 35 integrations per exposure. A total of two dither positions were obtained, resulting in 70 integrations and a total exposure time of 6988\,s.

The data were reduced using the standard \textit{JWST} Calibration Pipeline with the appropriate CRDS context. Stage~1 processing applied detector-level corrections, including reference pixel subtraction, non-linearity correction, dark current subtraction, and ramp fitting. Stage~2 applied flat-fielding, wavelength calibration, flux calibration, and WCS assignment to produce calibrated (\texttt{\_cal}) products. Stage~3 combined the dithered exposures and performed background subtraction and outlier rejection to generate the final two-dimensional rectified spectra. One-dimensional spectra were extracted using an aperture centered on the trace, with aperture corrections applied based on the wavelength-dependent MIRI LRS PSF.

\section{Analysis}
\label{sec:analysis_results}

\subsection{TDE Astrometry}

We performed astrometric calibration on the \textit{JWST} MIRI image using \textit{Gaia} DR3 \citep{Gaia_DR3_2023} positions of stars within the field to refine the absolute World Coordinate System (WCS) solution. The initial WCS solution was found to be offset by $0.11''$ in RA and $0.09''$ in Dec, and was corrected accordingly.

Using the astrometrically calibrated MIRI image, the position of the TDE was measured to be RA, Dec (J2000) = 01:46:55.4000, $-15$:22:15.870 in the ICRS frame. The centroid of the host galaxy nucleus is located at 01:46:56.0308, $-15$:22:14.837. The angular separation between the TDE and the host nucleus is $9.32 \pm 0.02''$, corresponding to projected physical separation of $9.08 \pm 0.02$~kpc.

\subsection{Optical and IR spectroscopy}

    \begin{figure*}[tbp]
        \centering
    	\includegraphics[width=\linewidth]{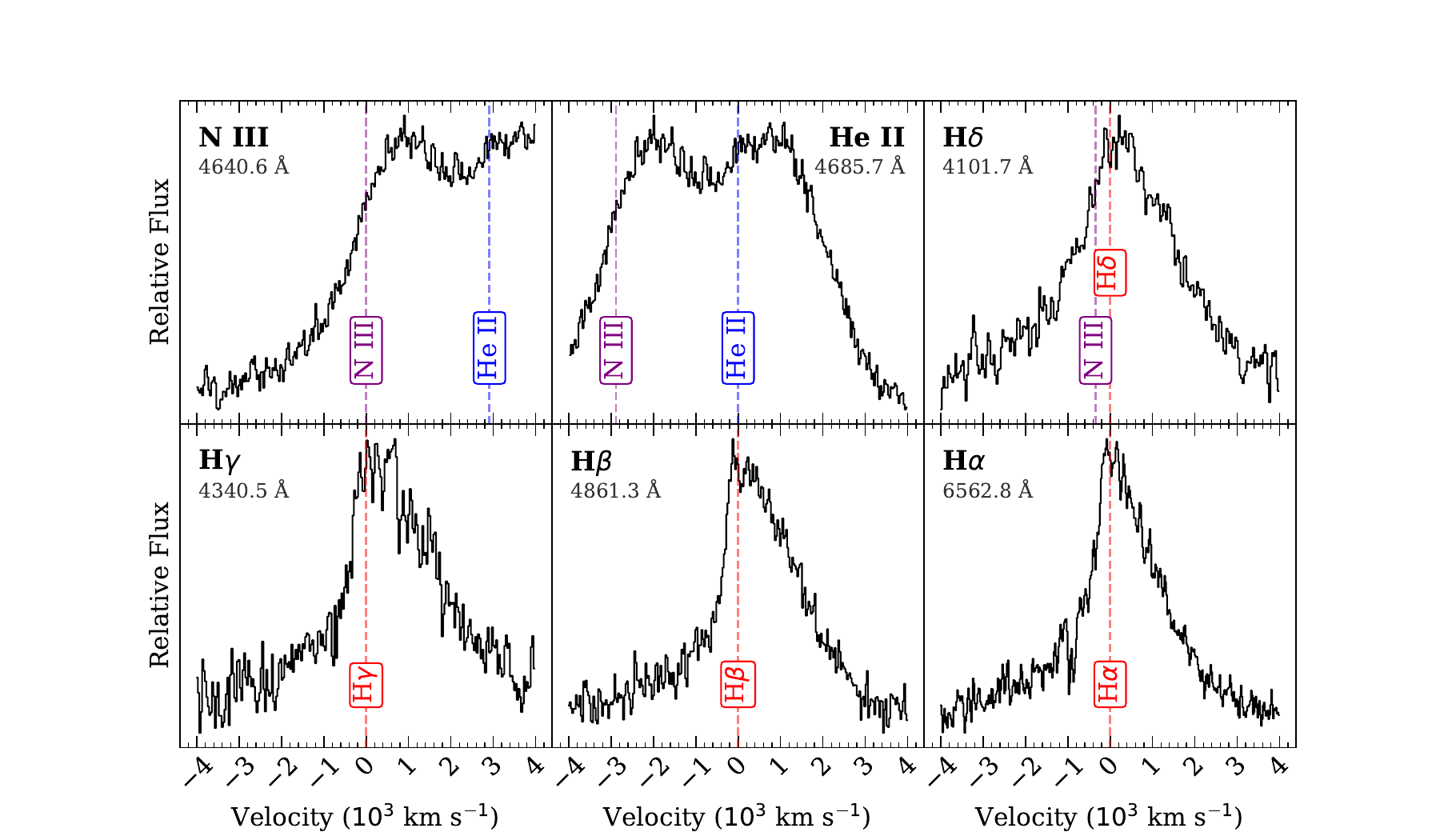}
    	 \caption{Optical emission-line profiles from the Keck/KCWI spectrum of TDE~2025abcr obtained at phase +29 days, shown in velocity space at
       the native instrumental sampling. The Balmer lines are asymmetric but are centered at the host-galaxy rest frame, with no
      significant velocity offset. The \ion{N}{3}+\ion{He}{2} complex, however, is clearly redshifted by \(\sim 1000\)~km~s\(^{-1}\).}
        \label{fig:kcwi_line_vels}
    \end{figure*}

    \begin{figure*}[tbp]
        \centering
    	\includegraphics[width=\linewidth]{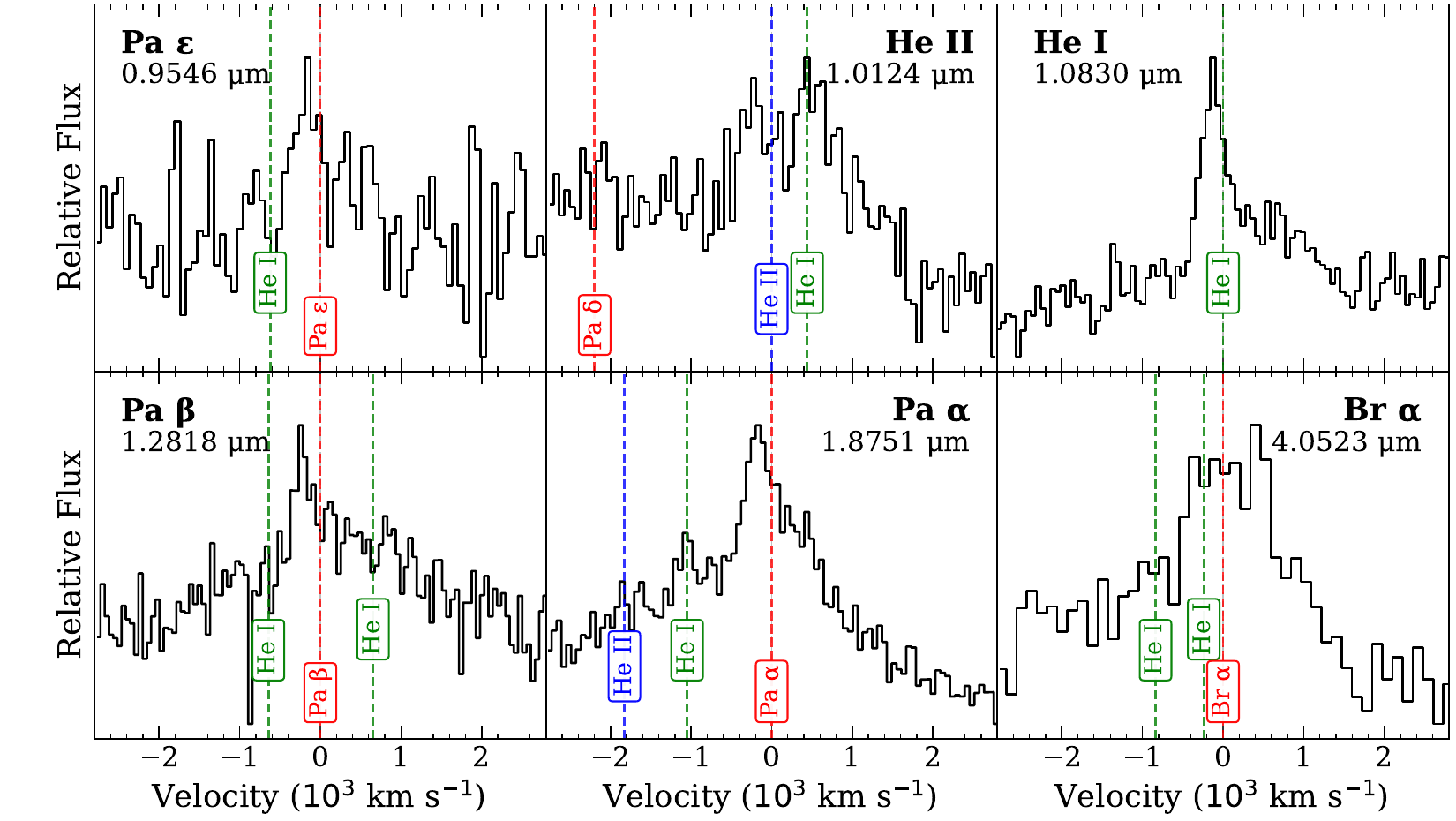}
    	\caption{\textit{JWST} IR emission lines observed at day \(+60\), plotted in velocity space at the native instrumental sampling. Several lines in the IR spectrum are blended. The profiles (particularly \ion{He}{1} \(1.083\,\mu\mathrm{m}\), Pa$\beta$ \(1.282\,\mu\mathrm{m}\), and Pa$\alpha$ \(1.875\,\mu\mathrm{m}\)) show extended red wings and peaks blueshifted on average by \(180\pm40~\mathrm{km\,s^{-1}}\).}
        \label{fig:jwst_line_vels}
    \end{figure*}

Both the optical (Figure \ref{fig:optical_spec}) and IR (Figure \ref{fig:jwst_spec}) spectra show broad H and He emission lines, consistent with the TDE-H-He spectroscopic classification. Prominent \ion{N}{3} emission (from Bowen fluorescence) is also detected, which requires a strong EUV/soft X-ray radiation field to resonantly pump \ion{He}{2} and subsequently excite \ion{N}{3} transitions \citep{Leloudas_etal_2019, Onori_2019_bowen, vanvelzen_2021_TDEclass}. The presence of Bowen fluorescence indicates efficient reprocessing in dense, likely CNO-enriched gas in the vicinity of the MBH. The IR spectrum contains a rich set of hydrogen recombination lines, including the Paschen, Brackett, Pfund, and Humphreys series, along with \ion{He}{1} and \ion{He}{2} emission.

Figures~\ref{fig:kcwi_line_vels} and \ref{fig:jwst_line_vels} show the major optical and IR emission lines in velocity space from our KCWI spectrum at day \(+29\) and our \textit{JWST} spectrum at day \(+60\), respectively. The H Balmer lines are centered at zero velocity with a FWHM of $\sim 3000~\mathrm{km~s^{-1}}$, narrower than the $\sim 10^4~\mathrm{km~s^{-1}}$ widths typical of the broader optical TDE population \citep[e.g.,][]{Hung_2017_tdelw, Holoien_2019, Nicholl_etal_2020, Charalampopoulos_2022}, but characteristic of the narrower-lined Bowen subclass \citep{Charalampopoulos_2022}.

The Balmer profiles are asymmetric, with extended red wings. The IR lines are more blended and have lower SNR than the optical lines. Nevertheless, they also show asymmetric red wings and slightly blueshifted peaks with a mean offset of \(-180\pm40~\mathrm{km\,s^{-1}}\), with respect to the host galaxy's rest frame. This is particularly evident in the \ion{He}{1} \(1.083\,\mu\mathrm{m}\),  Pa$\beta$ \(1.282\,\mu\mathrm{m}\), and Pa$\alpha$ \(1.875\,\mu\mathrm{m}\) lines. Such profiles can be interpreted as signatures of electron scattering in an optically thick medium, velocity gradients in an expanding or inflowing gas, or radiative transfer effects in a stratified outflow \citep[e.g.,][]{Roth_Kasen_2018, Parkinson_2022}.

\begin{figure*}
        \centering
    	\includegraphics[width=\linewidth]{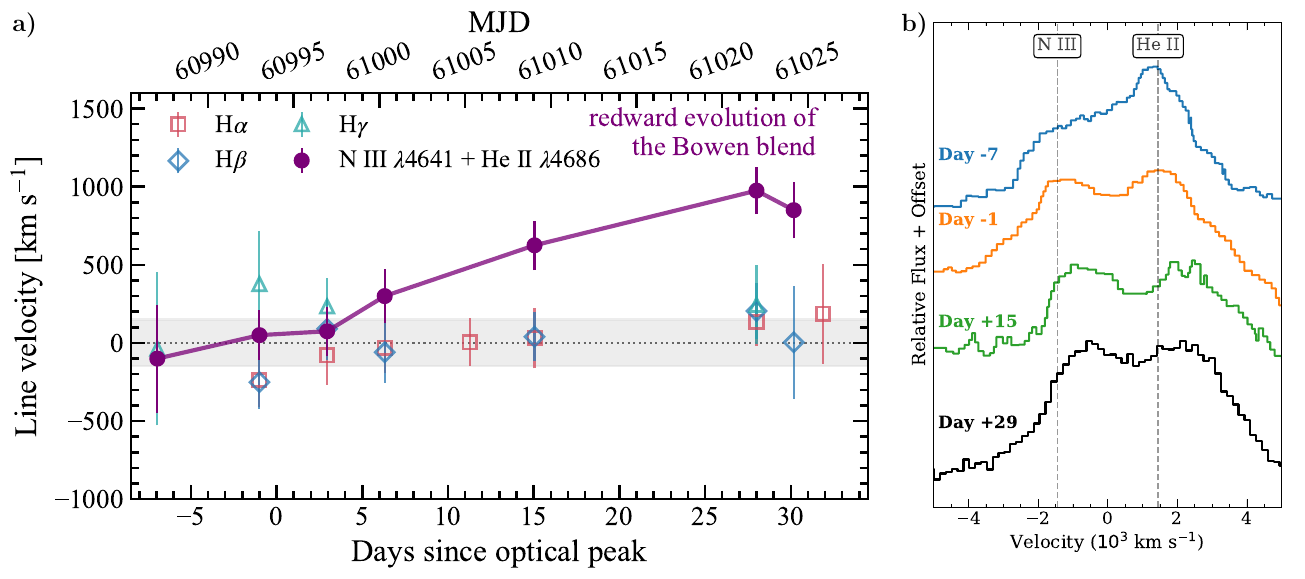}
        \caption{\textit{Left:} Time evolution of the velocities of prominent optical emission lines in TDE\,2025abcr. The Balmer lines show no significant velocity evolution, whereas the Bowen \ion{N}{3}+\ion{He}{2} complex exhibits a clear redward drift. The shaded gray region represents the systematic uncertainty floor of velocity measurements, accounting for modeling and wavelength-calibration uncertainties. \textit{Right:} Evolution of the \ion{N}{3}+\ion{He}{2} emission complex in velocity space. The top three spectra are taken from \citet{2026arXiv260210180S}; for clarity, only the smoothed spectra are shown. The bottom spectrum, obtained at day $+29$, is from our KCWI observation.}
        \label{fig:line_evol}
    \end{figure*}

An interesting feature of the optical spectra is the behavior of the Bowen \ion{N}{3} $\lambda4641$ + \ion{He}{2} $\lambda4686$ complex, which exhibits a clear, monotonic velocity evolution, whereas the H and He line velocities remain unchanged. In Figure \ref{fig:line_evol} (left panel), we show the velocity evolution of the Bowen complex alongside that of H Balmer lines in the optical spectral series presented by \citet{2026arXiv260210180S}, as well as in our KCWI spectrum. We note that the optical He lines were not sufficiently prominent to yield reliable velocity measurements. The continuum-subtracted Balmer lines were modeled with single Gaussians, while the Bowen complex was modeled with two Gaussians sharing a common velocity and width but having independent amplitudes. Uncertainties in the measured velocities were determined from the bootstrap scatter, with a conservative systematic floor of $150~\rm km~s^{-1}$ (represented by the shaded region) to account for unknown model and wavelength-calibration uncertainties in the spectral series of \citet{2026arXiv260210180S}. Despite this generous allowance for systematic uncertainty, the redward drift of the Bowen complex is readily apparent relative to the unchanging H-line velocities. For reference, the velocity resolutions of the various spectra in \citet{2026arXiv260210180S} ranged from $\sim 100$ to $350~\rm km~s^{-1}$, significantly smaller than the observed redward drift of the Bowen complex.

The right panel of Figure \ref{fig:line_evol} shows the evolution of the Bowen complex in velocity space. 
We observe that at early times (day $-7$), the feature is modestly blueshifted by $\sim 100~\mathrm{km~s^{-1}}$, while at later epochs it shifts progressively to the red, reaching $\sim 1000~\mathrm{km~s^{-1}}$ by day $+29$ in our Keck spectrum (Figure~\ref{fig:line_evol}). This monotonic drift is not seen in the H Balmer lines, indicating that the Bowen-emitting region is kinematically distinct from the  hydrogen-emitting gas. Additionally, the \ion{N}{3} component strengthens relative to \ion{He}{2} over time. Although a full radiative transfer analysis of the TDE emission region is beyond the scope of this work, we discuss possible interpretations in the context of reprocessed TDE emission in Section~\ref{sec:discussion}.

\subsection{SED modeling}
\label{sec:sed_modeling}

    \begin{figure*}[tbp]
        \centering
    	\includegraphics[width=\linewidth]{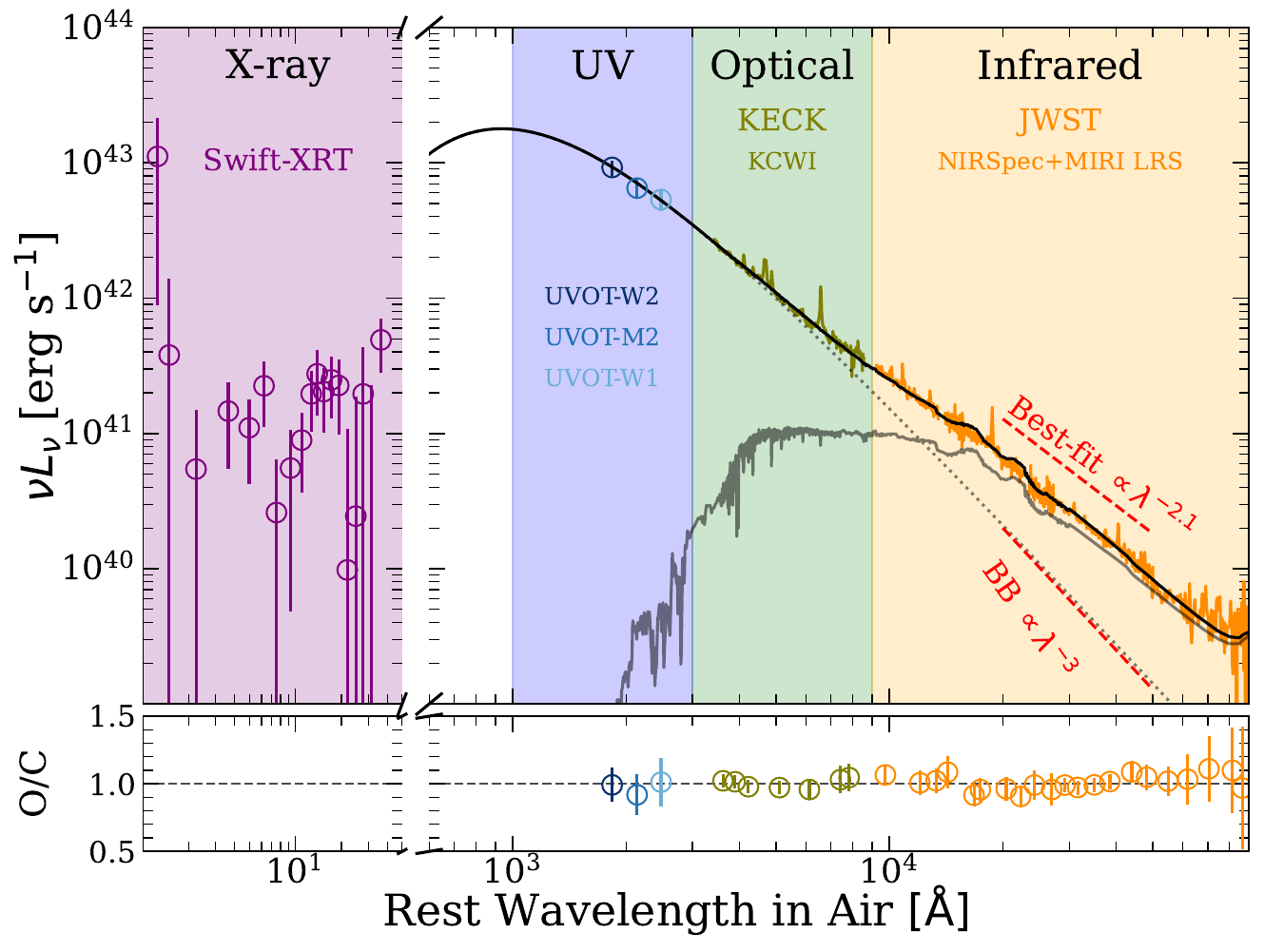}
    	\caption{X-ray to IR SED of TDE~2025abcr. The best-fit blackbody model to the UV--optical--IR data is shown as the dotted curve, the best-fit star cluster spectrum in gray, and the combined best-fit model as the solid black curve. The best-fit IR slope is indicated, along with the Rayleigh--Jeans slope expected for a blackbody. For completeness, the full KCWI (divided by $f_{\rm KCWI}$ for presentation) and \textit{JWST} spectra are shown in the upper panel; however, the emission lines were masked and the spectra binned for the SED fitting, as shown in the residuals panel (bottom). The X-ray data were not included in the SED fitting.}
        \label{fig:sed}
    \end{figure*}

We constructed a multi-wavelength SED of TDE~2025abcr using \textit{Swift}/UVOT observations obtained on 2026 January 16 (day $+62$) in the $UVW2$, $UVM2$, and $UVW1$ bands, together with the \textit{JWST} NIRSpec+MIRI spectrum obtained on 2026-Jan-14 (day $+60$), and the KCWI optical spectrum obtained on 2025-Dec-13 (day $+29$). The resulting SED is shown in Figure~\ref{fig:sed}. For completeness, we also show the stacked X-ray spectrum, but do not include it in  the SED analysis because of the uncertain origin of the X-ray emission. 

All UV, optical, and IR data were shifted to the rest frame in both wavelength and flux using the known redshift of the host galaxy. 
Wavelengths in air were used consistently for all data sources. The flux densities were converted to luminosity units as $\nu L_\nu$ in erg~s$^{-1}$. For the KCWI and \textit{JWST} spectra, only line-free continuum regions were included in the SED fit. Around each identified emission line, we masked pixels within \(\pm2\) times the observed line FWHM to exclude the line cores and wings. We combined the remaining continuum into broad wavelength bins---each much larger than the instrumental resolution element---
to average over noise and obtain continuum measurements for the SED. We assessed the sensitivity of the inferred IR slope to both the binning and line masking by repeating the fit with alternative bin widths and masks extending to \(\pm1\), \(\pm2\), and \(\pm3\) times the observed line FWHM; the resulting slopes were consistent within \(1\sigma\).
The continuum measurements used in the fit are shown in the lower panel of Figure~\ref{fig:sed}.

Figure~\ref{fig:sed} shows immediately that the IR continuum deviates strongly from the Rayleigh--Jeans tail of the UV--optical blackbody.  A direct power-law fit gives
\[
\nu L_\nu \propto \lambda^{-2.13 \pm 0.04},
\]
substantially shallower than the blackbody expectation, $\nu L_\nu\propto\lambda^{-3}$.  The excess IR emission could be explained by wavelength-dependent free--free thermalization in the TDE reprocessing layer, an unresolved stellar population at the transient position, or, less plausibly given the observed spectral shape, from hot dust. These alternatives are examined in Section~\ref{sec:discussion}.

The stellar component of the model is physically motivated for two reasons: (1) a wandering MBH likely requires a surviving stellar cluster to efficiently refill its loss cone and thereby sustain an observable TDE rate; and (2) cool main-sequence stars and evolved giants within such a cluster can enhance the long-wavelength tail because their integrated emission contributes strongly at optical and near-IR wavelengths. Increasing age generally raises the optical/IR mass-to-light ratio and reddens the continuum, while increasing metallicity strengthens line blanketing at short wavelengths and shifts the integrated light toward cooler stellar temperatures. 

If real, this compact stellar population has a natural dynamical interpretation: it may be the surviving nucleus or dense core of the satellite galaxy that delivered the wandering MBH during a minor merger.  We return to this connection in Sections~\ref{sec:discussion} and \ref{subsec:origin_of_MBH}.

Motivated by the viable stellar interpretation, we modeled the SED jointly with two components: a single-temperature blackbody representing the TDE continuum, and a stellar component representing an unresolved stellar cluster at the TDE location. A single-temperature blackbody is a reasonable approximation at this phase, when the TDE flux is still declining and has not yet transitioned to a late-time plateau \citep[e.g.,][]{Gezari_2021_TDE_Rev, vanvelzen21_TDE_tdes, Hammerstein_2023_TDE, Yao_Yuhan_2023_TDEs}. 
The stellar component was modeled using a precomputed grid of Flexible Stellar Population Synthesis (\texttt{FSPS}; \citealp{Conroy_Gunn_2010_FSPS}) templates spanning ages from 0 to 14 Gyr and metallicities $\log (Z/Z_\odot)$ from $-2$ to $+0.5$, assuming a single stellar population and a \citet{2001_Kroupa} initial mass function. To reduce compute time, we interpolated within this grid instead of calling \texttt{FSPS} at each likelihood evaluation.

The free parameters of the model are the stellar mass of the unresolved cluster, $M_{\rm NSC}$, its age, its metallicity $\log(Z/Z_\odot)$, the TDE blackbody radius $\log r {\rm [cm]}$, and the blackbody temperature $\log T {\rm [K]}$. Because the KCWI spectrum was obtained substantially earlier than the \textit{Swift} and \textit{JWST} observations, we introduced an additional multiplicative factor, $f_{\rm KCWI}$, to scale the KCWI continuum relative to the day $+60$ SED. This effectively assumes that the SED shape does not evolve strongly between days $+29$ and $+60$. That assumption is supported by the weak rate of temperature change reported by \citet{2026arXiv260210180S}. Moreover, using the \textit{Swift}/UVOT $UVW2-UVW1$ color, we find no significant color evolution: the color changes from $-0.23 \pm 0.14$ mag at day +28 to $-0.25 \pm 0.24$ mag at day +62, corresponding to a difference of $-0.02 \pm 0.28$ mag---consistent with no color evolution. The KCWI spectrum plotted in Figure~\ref{fig:sed} has been divided by $f_{\rm KCWI}$ for presentation. We also included a nuisance parameter, $\ln S_0$, to account for any additional unexplained scatter beyond the formal measurement uncertainties.

Posterior distributions were sampled using \texttt{emcee} \citep{Foreman-Mackey_etal_2013_emcee} with 48 walkers and $10^5$ steps, discarding the first 30\% of the chains as burn-in. Convergence was evaluated by monitoring trace plots for stability and calculating the integrated autocorrelation time ($\tau$). We ensured that the total chain length exceeded $50\tau$ for all parameters. The best-fit model is shown in Figure~\ref{fig:sed}, and the posterior distributions are shown in Figure~\ref{fig:sed_corner} in the Appendix.

The posterior distributions for the fitted parameters are as follows: blackbody temperature $\log T~[{\rm K}] = 4.59^{+0.06}_{-0.05}$,   blackbody radius $\log r~[{\rm cm}] = 14.10 \pm 0.05$, KCWI scaling factor $f_{\rm KCWI} = 1.72^{+0.13}_{-0.12}$. 
For the unresolved stellar component, we infer $\log (M_{\rm NSC}/M_\odot) = 7.57 \pm 0.02$, an age of $0.71^{+0.38}_{-0.45}$ Gyr, and a metallicity of $\log (Z/Z_\odot) = 0.40^{+0.07}_{-0.11}$. 
We evaluate these parameters and discuss their implications in Section \ref{sec:discussion}.

\subsection{Host galaxy spectra and properties}
\label{sec:host_galaxy}

\begin{figure*}[tbp]
    \centering
    \includegraphics[width=\linewidth]{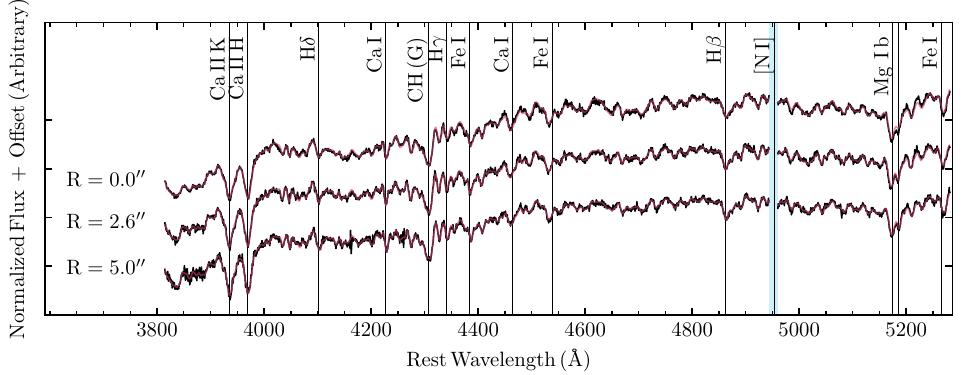}
    \caption{Representative Keck KCWI spectra (black) and LOSVD-broadened stellar template fits (red) from three spatial bins located at increasing radius (top down) in the host galaxy. No significant emission-line contribution is evident in the spectra. Several stellar absorption features are marked. The 5200~\AA\,\ion{N}{1} sky line (vertical blue band) is masked in the spectral fits.} 
    \label{fig:host_ppxf}
\end{figure*}

We extracted spatially-resolved spectra for the host galaxy from our KCWI observations described in Section~\ref{sec:obs}. To reduce contamination from the TDE, we masked the region within $1''$ of the TDE centroid and then voronoi-binned the data to uniform SNR of ${\sim}60$. For each of the 68 resultant spatial bins, we performed spectral fitting on the data using the penalized pixel-fitting method \citep[\texttt{pPXF};][]{Cappellari_2012_ppxfsoft, Cappellari_2017_ppxfpaper}, which models each observed spectrum as a linear combination of stellar template spectra convolved with a parameterized line-of-sight velocity distribution (LOSVD). 
Aside from features noted here, the \texttt{pPXF} configuration used here is identical to that in \cite{liepoldetal2025}. 

For the stellar templates, we selected a subset of the 1273-template Indo-US library comprising of 526 stellar templates where \citet{valdesetal2004} presents well-determined spectral types in their Table~2 and measurements $T_{\mathrm{eff}}$, $\log(g)$ and ${\rm [Fe/H]}$ in their Table~3. The spectral resolution of the Indo-US library is $\sigma_{inst} =0.57\,\text{\AA}\pm0.03\,\text{\AA}$ \citep{valdesetal2004,beifiorietal2011}. In comparison, we determined the line spread function (LSF) of the blue-side of the KCWI small slicer (from arc-lamp exposures) to be well approximated by a Gaussian with $\sigma_{\rm{inst}} = 0.56\,\text{\AA} \approx 35~\mathrm{km~s^{-1}}$, or $1.32\,\text{\AA}\approx 82~\mathrm{km~s^{-1}}$ FWHM. 
Because of the close match in spectral resolution between the templates and KCWI, no broadening to the templates is needed to match KCWI's LSF.

We performed \texttt{pPXF} fitting on each spatial bin to determine the stellar LOSVD.
We fit for the first two moments of the LOSVD: radial velocity and the velocity dispersion. The best-fitting LOSVD provides the stellar radial velocity $V$ and velocity dispersion $\sigma$; 
Figure~\ref{fig:host_ppxf} shows
the KCWI galaxy spectra and the LOSVD-broadened stellar template fits for three representative spatial bins at increasing radius from the galaxy center.
The spectra show the typical absorption features of early-type galaxies and no significant nebular emission lines in any of the spatial bins. 
From the light-weighted combination of templates, we estimate the metallicity  averaged over the KCWI field of view of the host galaxy is ${\rm [Fe/H]} \approx +0.12$.

We will present detailed properties of the spatially-resolved stellar kinematics and the dynamically measured mass for the central black hole in the host galaxy from triaxial orbit modeling \citep{quennevilleetal2022,liepoldetal2023} in a separate paper.
Here, we use the $M_\bullet$--$\sigma$ scaling relation and the host galaxy's stellar velocity dispersion $\sigma$ 
to obtain a crude estimate of the black hole mass $M_\bullet$.
We find $\sigma$ to rise sharply towards the center of the host galaxy, so 
the value of $\sigma$ used to estimate $M_\bullet$ depends on the aperture over which the spatially resolved $\sigma$ is averaged (see, e.g., discussion in Sec.~2 of \citealt{McConnell_2013_Msigma}).
We perform a luminosity-average of $\sigma$ from our KCWI measurements and find $\langle\sigma\rangle =224\pm4~\mathrm{km~s^{-1}}$ within the central radius $R<0.5''$, $\langle\sigma\rangle= 199\pm9~\mathrm{km~s^{-1}}$ for $0.5''<R< 7''$ (where $7''$ is the outer reach of our KCWI coverage), and $\langle\sigma\rangle= 205\pm14~\mathrm{km~s^{-1}}$ when luminosity-averaged over all KCWI bins.

Using the $M_\bullet$--$\sigma$ relation of \cite{McConnell_2013_Msigma} and the luminosity-weighted dispersion across all KCWI spatial bins, $\langle\sigma\rangle=205\pm14~\mathrm{km~s^{-1}}$, which samples a radius roughly comparable to the galaxy's estimated effective radius based on the early-type galaxy mass--size relation \citep{Shen_2003_gal_size_mass}, we infer a \textit{nuclear} black hole mass of $\log_{10}(M_\bullet/M_\odot)=8.38\pm0.42$.
This uncertainty includes the intrinsic scatter of \(0.37\) dex in the \(M_{\bullet}\)--\(\sigma\) relation, added in quadrature with the measurement uncertainty.
We note that the velocity dispersion reported by \citet{2026arXiv260210180S}, $\sigma = 324 \pm 14~{\rm km\,s^{-1}}$, and the corresponding black hole mass, $\log(M_{\bullet}/M_{\odot}) = 9.41 \pm 0.31$, are substantially higher than our estimates.  We have performed a number of tests to assess the robustness of our measured velocity dispersions by varying, e.g., the number of velocity moments included in the LOSVD, the multiplicative polynomial order used to model the continuum shape, the choice of Indo-US templates, and the spectral window used for fitting. All of these tests show
negligible ($<5~\mathrm{km~s^{-1}}$) impact on our measurements.

The host galaxy PGC~174145 lies in the 2MASS XSC footprint \citep{jarrettetal2000}, where it was found to have a K-band magnitude (\texttt{k\_m\_ext}) of 11.583. With extinction correction this yields $M_K =-25.175$, and application of Equation~(1) of \citet{liepoldma2024} suggests a total stellar mass of $\log(M_* / M_\odot) \approx 11.52$. PGC~174145 also lies in the WISE survey footprint \citep{wrightetal2010}, where it has $W1 = 12.26$ (\texttt{w1mpro}) and $W2 = 12.293$ (\texttt{w2mpro}). 
Following the stellar mass calibration in \citet{cluveretal2014}, this suggests $\log( M_* / M_\odot) \approx 11.11$, a factor of 2.5 lower than the K-band-based estimate.

The K-band calibration from \citet{liepoldma2024} correlated K-band absolute magnitudes against stellar masses measured from dynamical models and stellar population models with variable IMFs. When comparing against prior galaxy stellar mass functions derived from stellar population models with Milky-Way-like IMFs, that work found mass scales which were larger by a factor similar to the IMF mismatch parameters measured for their galaxy sample, ${\sim}1.84$. The stellar mass calibration in \citet{cluveretal2014} relied on stellar masses from GAMA \citep{tayloretal2011}, which also used a Milky-way-like Chabrier IMF.

In Legacy Survey DR10 imaging, the host galaxy appears to be a morphologically regular elliptical. Accordingly, we can interpret the stellar mass as a bulge mass and apply the $M_\bullet$--$M_{\rm{bulge}}$ relation from \citet{McConnell_2013_Msigma} to infer $M_{\bullet}$. We estimate $\log(M_\bullet / M_\odot) \approx 9.01 \pm 0.35$ when using the K-band-based stellar mass estimate and $8.58 \pm 0.35$ when using the WISE magnitudes, a factor of 2--5 above estimates from the $M_\bullet$--$\sigma$ relation. 

The host-galaxy analysis shows that TDE~2025abcr occurred in the outskirts of a massive ($\log(M_* / M_\odot) \approx$ 11.1--11.5), quiescent, early-type galaxy, with no signs of an active galactic nucleus (AGN), such as broad H Balmer lines, narrow emission lines, or WISE colors indicative of AGN activity. The galaxy likely harbors a supermassive black hole (SMBH) of mass $\log(M_\bullet / M_\odot) \approx$ 8.3--9.0. These conclusions---a massive early-type host, no spectroscopic or WISE evidence for an AGN, and a very massive central SMBH---broadly agree with the independent host analysis of \citet{2026arXiv260210180S}.

\section{Discussion}
\label{sec:discussion}

\subsection{Infrared excess}

    \begin{figure}[!t]
        \centering
    	\includegraphics[width=\linewidth]{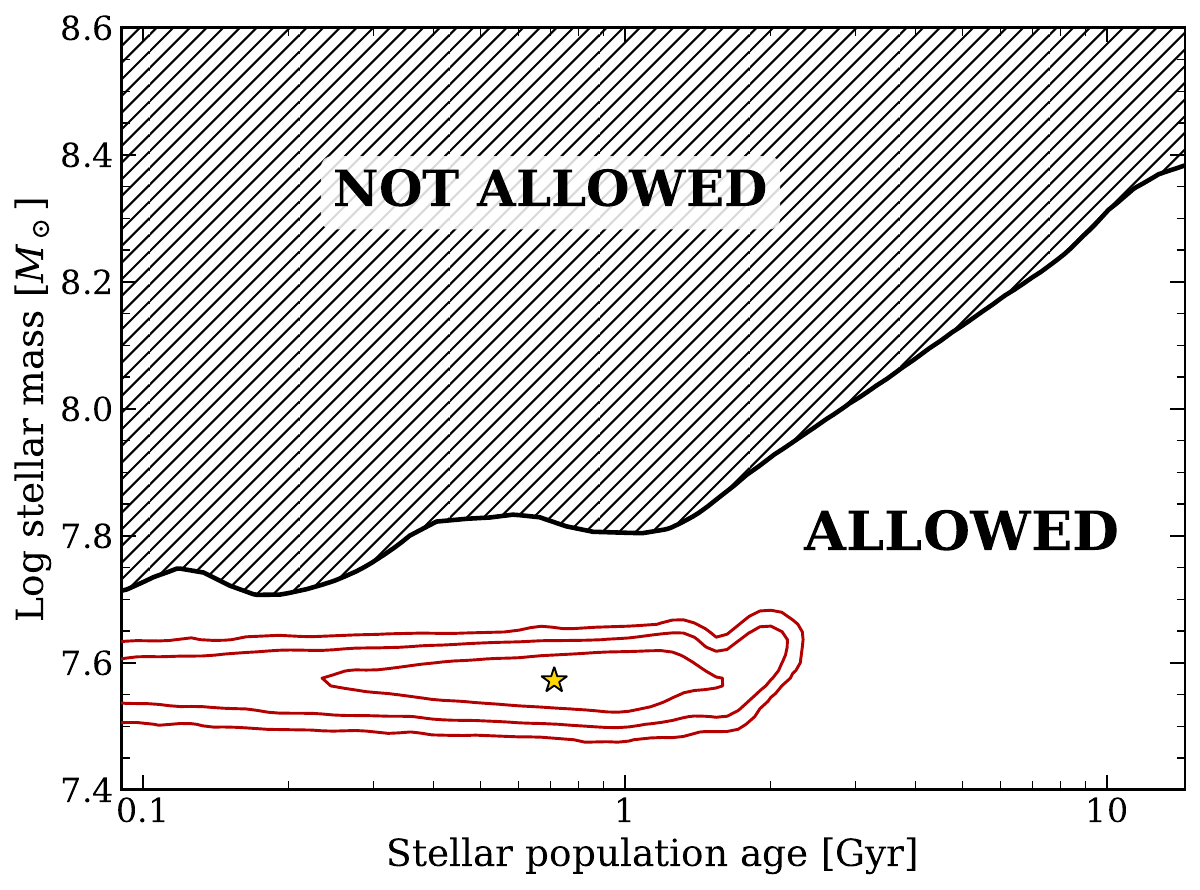}
    	\caption{Stellar population constraints in age–mass space. The hatched region is ruled out by the observations: a stellar population in this region would overpredict the flux in at least one observed UV, optical, or IR data point by $>5\sigma$. The unshaded region remains consistent with the data. The red contours show the MCMC posterior from the full SED fit in age–mass space (Section \ref{sec:sed_modeling}), and the star represents the median of the posterior distribution. The boundary between the allowed and excluded regions is obtained after profiling over metallicity, so at each age and mass it reflects the most conservative constraint.}
        \vspace{2pt}
        \label{fig:age_mass}
    \end{figure}

There are three possible origins for the IR excess observed in the SED of TDE~2025abcr: (1) thermal emission from dust heated by the TDE, (2) an unresolved stellar component, or (3) free--free emission from ionized gas in a reprocessing gas layer or outflow.

\begin{itemize}
\item A dust echo is disfavored by the observed SED shape: thermal dust emission near the sublimation temperature ($\lesssim 1500$\,K) would be expected to peak at wavelengths $\gtrsim 2~\mu$m and to decline more steeply blueward of that peak, unlike the observed IR continuum of TDE~2025abcr \citep{Jiang_2016_dustecho, vanvelzen_2016_TDedust, Jiang_2021_IRechoes, vanvelzen_2021_TDEclass, 2023ApJ...959L..19D}. We therefore rule out dust as the dominant source of the IR excess.

\item The second possibility is that the IR excess comes from an unresolved stellar component, as illustrated in Figure~\ref{fig:sed}. If interpreted as starlight, the inferred stellar mass ($\log M_{\rm NSC}/M_\odot = 7.57 \pm 0.02$) is large for a typical NSC and would more likely suggest the compact core of a stripped satellite galaxy \citep{Boker_2004_NSC, Seth_Anil2014, Neumayer_2020_NSCrev}. 

\citet{2026arXiv260210180S} likewise determined that any underlying stellar component at the position of TDE~2025abcr must be low mass, deriving an upper limit of \(\log M_\star/M_\odot < 7.4\) by assuming a similar  \(g\)-band mass-to-light ratio ($M/L$) as EP240222A \citep{Jin_2025_EP24}. However, that limit is necessarily age dependent, because the stellar $M/L$ changes with age. Older stellar populations have higher $M/L$, meaning that for the same observed luminosity they can be more massive than younger populations. So the inferred stellar-mass upper limit is best treated as a function of age rather than as a single number. This is the approach we adopt in Figure~\ref{fig:age_mass}. For each point in age--mass space, we ask whether a stellar population would overproduce any observed UV, optical, or IR flux point after profiling over metallicity. The resulting exclusion boundary therefore gives the most conservative upper limit on stellar mass as a function of age. We find that for young populations (\(\lesssim 1\) Gyr), stellar masses above \(\log M_\star/M_\odot \approx 7.8\) are excluded, whereas for older populations the limit becomes progressively weaker, reaching \(\log M_\star/M_\odot \approx 8.2\)--8.4 by \(\sim 10\) Gyr. The median solution, marked in Figure~\ref{fig:age_mass} by a star, lies safely within the allowed region and favors a young stellar population ($\lesssim 2$ Gyr).

Such a young satellite population contrasts with the inferred old $\sim 9$~Gyr age of the early-type host galaxy \citep{2026arXiv260210180S}. However, satellite populations need not closely track the stellar populations of their central galaxies. The Satellites
Around Galactic Analogs (SAGA) Survey analyzed 378 satellite galaxies and found no significant correlation between satellite quenching and host color or host star formation rate \citep{Geha_2024_SAGA}. 
SAGA also found that quenching proceeds sufficiently slowly that star-forming satellites persist over a wide range of distance from the central galaxy, and that the quenched fraction rises toward the inner \(\sim 100\) kpc rather than being determined simply by whether the central galaxy is old or passive. 

These results support a scenario in which the stellar component associated with TDE~2025abcr is the remnant core of a low-mass satellite that continued forming stars until relatively recently and was quenched only within the last 1--2~Gyr, for example during infall into the halo of the giant elliptical host. This interpretation is also relevant to the preferential association of TDEs with quenched and post-starburst systems, and to the population of low-mass quenched galaxies hosting undermassive SMBHs identified by \citet{2026ApJ...998L..25R}. This is also broadly consistent with recent work on the TDE delay-time distribution, which finds that the TDE rate peaks at $\sim 1$~Gyr after the most recent burst of star formation \citep{Shepherd_2026_TDE_DTD}. In other words, a stellar population of age $\sim 1$~Gyr is the most likely to host a TDE at the time of observation.

\item The third possibility is that the IR excess is intrinsic to the TDE, arising either from free--free emission within a dense, stratified reprocessing layer/outflow or from viewing-angle effects. The free--free absorption opacity is frequency dependent, with $\alpha_{\rm ff} \propto \lambda^{2}$. This can produce an IR continuum that is shallower than the Rayleigh--Jeans tail of a single blackbody, because photons at different wavelengths thermalize at different depths and therefore emerge from regions with different temperatures and physical conditions. The observed continuum therefore reflects radiative transfer through an extended, scattering-dominated medium rather than emission from a single thermal photosphere \citep[e.g.,][]{Roth_2016_TDEs}.

In the \citet{Roth_2016_TDEs} framework, the continuum slope depends on the coupled structure of density, temperature, ionization, opacity, and geometry, as well as on where photons thermalize in the flow. Therefore, the observed IR slope cannot be uniquely inverted into a photospheric density law without a self-consistent radiative-transfer model. This is particularly important if the optical and IR emission originates from different regions, since the analytic mapping between spectral slope and density assumes a continuous atmosphere with a smooth thermalization structure.

As an illustrative estimate, if one applies Equation~(6) of \citet{Margutti_2019} to the measured IR slope, $\nu L_\nu \propto \lambda^{-2.13}$, one obtains a formally steep density profile, $\rho \propto r^{-6.13}$, using $n=(2\alpha-8)/(3\alpha-7)$ with $\alpha=2.13$, which would indicate a wind with a strongly decreasing mass-loss rate. However, this inference is highly model-dependent. Such a steep decline would imply a sharp truncation of the photospheric layers, with the optical depth dropping over a narrow radial range, approaching an exponential cutoff. It is unclear whether this is consistent with the extended, continuously stratified, and time-dependent outflows expected in TDEs \citep{Roth_2020_review}.

In addition, the IR slope may be affected by the temperature and ionization structure of the flow, by deviations from spherical symmetry, and by viewing-angle effects in an anisotropic reprocessing layer. In particular, radiation hydrodynamical simulations show that similar broad spectral slopes, comparable to the one observed for TDE~2025abcr, can arise from the same underlying system viewed at different inclinations, consistent with this event being observed significantly off-axis \citep[e.g.,][]{2022_Thomsen}. Thus, while the observed IR continuum is broadly consistent with emission from an extended reprocessing medium, inferring a unique photospheric structure from the slope is not straightforward and likely requires detailed time-dependent radiative-transfer calculations.

\item Intrinsic TDE emission (free--free or viewing-angle effects) currently provides the most plausible explanation for the IR excess since it is a natural prediction of reprocessed emission in TDEs \citep[e.g.,][]{Roth_2016_TDEs, Lu_Bonnerot_2020, Thomsen_2022_TDEs}, and has been interpreted as such for early IR emission from TDE~2019azh \citep{Reynolds_2025_TDE_IR}. The comparison is especially close in the continuum phenomenology: both events show an early IR excess above the Rayleigh--Jeans tail that cannot be fit self-consistently as hot dust across the available near-/mid-IR wavelengths, and both favor wavelength-dependent free--free opacity in a dense reprocessing envelope.  TDE~2019azh was followed from before peak to $>1750$~days: its early near-IR flux rose and faded within the first few tens of days, whereas emission after $\gtrsim200$~days was consistent with a cooler, distant dust echo.  TDE~2025abcr currently has only the day $+60$ IR spectrum, so its temporal evolution is not yet constrained.  Its asymmetric H/He profiles and evolving Bowen complex provide complementary spectroscopic evidence that the reprocessing gas is changing.

Furthermore, unlike the intrinsic TDE-emission interpretation, the stellar-cluster scenario requires specific conditions to reproduce the observed IR slope. The mass--age constraints require a cluster mass of $M_{\rm NSC}\approx10^{7.6}\,M_{\odot}$ and an age of ${<}2$ Gyr. We performed a grid search in age--metallicity space using \texttt{FSPS} stellar-population models to determine whether the observed IR slope, $-2.13\pm0.04$, arises generically or requires particular population parameters. The results, presented in Figure~\ref{fig:fsps_slope} in the Appendix, show that the observed slope is reproduced only within a relatively narrow, age-dependent band of super-solar metallicities, rather than across a broad range of stellar populations.

However, the presence of a young stellar cluster at the TDE location cannot currently be ruled out. A straightforward way to distinguish between these possibilities is to obtain additional spectroscopic data at a later epoch, once the TDE has settled onto its UV--optical plateau. By that stage, the reprocessing layer should have weakened substantially, and any associated free--free emission is expected to decline. However, if the IR excess originates from an underlying stellar cluster, the IR flux should remain approximately at its current level. 
\end{itemize}

\subsection{Evidence for an Evolving Reprocessing Layer}
\label{sec:evolving_reprocessing}

The spectroscopic evolution of TDE~2025abcr suggests that the line-forming and reprocessing region is evolving rapidly on a timescale of weeks. The key observational facts are as follows:

\begin{enumerate}
    \item The \ion{N}{3} $\lambda 4641$ + \ion{He}{2} $\lambda 4686$ complex shows a clear velocity evolution, shifting from $\sim -100~\mathrm{km~s^{-1}}$ at day $-7$ to $\sim +1000~\mathrm{km~s^{-1}}$ by day $+29$.
    \item The \ion{N}{3} component strengthens relative to \ion{He}{2} over the same period.
    \item The H Balmer lines on day $+29$ are asymmetric but do not show any velocity shift with respect to the galaxy's rest frame. The IR emission lines on day $+60$ are also asymmetric but do show a blueshift of $\sim 200~\mathrm{km~s^{-1}}$. 
    \item The X-ray flux declines over a similar timescale and becomes undetectable by $\sim$day $+40$ (assuming that the X-ray flux is associated with the TDE).
    \item The IR SED slope is shallower than a blackbody.
\end{enumerate}

\subsubsection{Line evolution and the inner reprocessing region}

The \ion{N}{3} + \ion{He}{2} complex likely originates in a denser and more compact region that is more directly coupled to the EUV/soft X-ray radiation field \citep{Leloudas_etal_2019,Onori_2019_bowen,Nicholl_etal_2020,Charalampopoulos_2022}. 
If the observed velocity drift of order $10^3~\mathrm{km~s^{-1}}$ were interpreted as Keplerian motion around a $\sim 10^{6}\,M_\odot$ MBH, the corresponding orbital timescale would be $\sim 30$\,yr, far longer than the few-week evolution observed in the \ion{N}{3} + \ion{He}{2} complex.
The monotonic velocity drift of the \ion{N}{3} + \ion{He}{2} complex therefore cannot be Keplerian orbital motion of some emitting ``clump'' of gas. A more natural interpretation is that the observed velocity shift reflects a changing view of a (possibly) asymmetric inner velocity field, driven by evolving optical depth, geometry, and illumination \citep{Roth_Kasen_2018,Nicholl_etal_2020, Parkinson_2022}. In this picture, the early-time blueshift may develop because the Bowen emission is weighted toward the near side of an asymmetric flow, while at later times the emitting surface shifts and the receding side contributes more strongly, producing the observed redward drift. 

The absence of comparable velocity evolution in the Balmer lines, together with their relatively narrow widths and stationary centroids, favors a more extended and/or symmetric hydrogen-emitting region that is less sensitive to changes in the inner ionizing source. However, optical-depth effects and projection within a stratified reprocessing layer could also produce narrow line cores without requiring the emitting gas to reside at a single radius. In either case, the hydrogen-emitting region appears distinct from the more compact, evolving Bowen-emitting zone.

The asymmetric extended red wings seen in the H Balmer lines on day \(+29\) and in the IR emission lines on day \(+60\) are consistent with formation in an electron-scattering-dominated photosphere, where repeated scatterings in an expanding reprocessing layer broaden the lines and produce a characteristic red wing. The blueshifted peaks observed in the IR lines also qualitatively match the predictions of \citet{Roth_Kasen_2018}, in which line formation in an expanding, optically thick outflow produces a blueshifted emission peak together with an asymmetric red wing. Similar profiles were previously observed in TDE~2019qiz \citep{Nicholl_etal_2020}, and a few others \citep[see][]{Charalampopoulos_2022}. The Balmer lines, however, do not show a similar blueshift, indicating evolving physical conditions within the outflow between day \(+29\) and day \(+60\).

The fading of the X-ray emission---if indeed associated with the TDE---also supports an evolving inner reprocessing region. Changes in the soft X-ray/EUV source, or increasing obscuration of the accretion flow by the reprocessing gas, should directly affect the ionization structure and the part of the flow that dominates the observed line emission. The X-ray fading could therefore trace either intrinsic weakening of the high-energy source or increasing obscuration by the evolving reprocessing layer, both consistent with rapid changes in the inner geometry of the system \citep{Roth_Kasen_2018,Nicholl_etal_2020, Stein_2021, Parkinson_2022, Malyali_2024,2026arXiv260210180S}. \citet{2026arXiv260210180S} independently proposed obscuration by optically thick stellar debris as an explanation for the rapidly fading soft X-ray component.

\subsubsection{Bowen fluorescence and ionization-state evolution}

Changes in density, optical depth, or covering fraction could also alter the efficiency of line formation and help explain the strengthening of \ion{N}{3} relative to \ion{He}{2}. 
The \ion{N}{3} $\lambda 4640$ line is produced through the \ion{He}{2} Ly$\alpha \rightarrow$ \ion{O}{3} $\rightarrow$ \ion{N}{3} resonance chain, and its luminosity depends on both the ionizing continuum and the local conditions in the line-forming gas \citep{Schachter_1989,Leloudas_etal_2019,Onori_2019_bowen}. The strengths of the \ion{O}{3} and \ion{N}{3} fluorescence lines can depend sensitively on the ionization balance of oxygen and nitrogen. From their ionization potentials, both \ion{O}{3} and \ion{N}{3} are favored in a radiation field with abundant photons in the $\sim 30$--$50$ eV range. If photons above $\sim 50$ eV are too abundant, however, oxygen and nitrogen will be driven to higher ionization states such as \ion{O}{4} and \ion{N}{1}V, thereby reducing the populations available to participate in the Bowen cascade and suppressing the \ion{N}{3} $\lambda 4640$ line relative to \ion{He}{2} $\lambda 4686$. 

In a multi-zone TDE outflow, the weak early-time \ion{N}{3} $\lambda 4640$/\ion{He}{2} $\lambda 4686$ ratio may reflect either an initially hard ionizing spectrum that over-ionizes O and N or emission weighted toward higher-ionization regions, while the subsequent strengthening of \ion{N}{3} $\lambda 4640$ may indicate that the ionizing continuum softens with time or that the emission becomes weighted toward regions with more favorable conditions for \ion{O}{3} and \ion{N}{3}. A useful observational test of this interpretation would be to observe the associated \ion{O}{3} Bowen lines, especially the strong near-UV feature around $3133$\,\AA\ and the optical lines near $3444$--$3750$\,\AA. If Bowen fluorescence is driving the observed evolution, these \ion{O}{3} lines should show the same qualitative trend as \ion{N}{3} $\lambda 4640$.

We therefore interpret the data as evidence for a rapidly evolving inner reprocessing region. More specifically, the coupled evolution of the Bowen complex and the fading X-ray emission is consistent with the emergence, buildup, or rapid reconfiguration of an asymmetric reprocessing layer surrounding the inner accretion flow. 
A full radiative-transfer treatment is beyond the scope of this work, but such modeling should provide a more definitive explanation for the observed line evolution.

\subsection{Origin of the wandering black hole}
\label{subsec:origin_of_MBH}

 \begin{figure}[!t]
        \centering
    	\includegraphics[width=\linewidth]{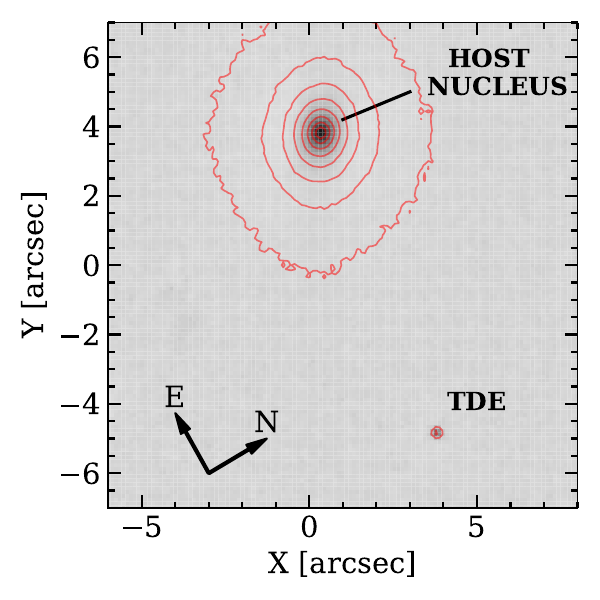}
    	\caption{Unsharp-masked \textit{JWST} MIRI F560W image. The galaxy emission has been largely subtracted out, leaving behind just the nucleus and the TDE. The red curves trace the flux contours of the galaxy for reference. No tidal tails or other large-scale structures indicative of a major galaxy merger are evident.}
        \label{fig:unsharp}
    \end{figure}

We estimate the mass of the MBH responsible for the TDE using \texttt{MOSFiT} \citep{Guillochon_2018_mosfit, 2019_Mockler}, finding $\log(M_{\bullet}/M_{\odot}) = 6.70 \pm 0.11$ (see Appendix for details). An independent estimate based on empirical scaling relations between black hole mass and TDE peak luminosity gives a value of $\log(M_{\bullet}/M_{\odot}) = 6.1 \pm 0.5$ \citep{Mummery_2024_BH-TDE_scaling, 2026arXiv260210180S}. Although neither method is perfect---and more reliable black hole mass estimates are now generally thought to come from late-time SED modeling with relativistic accretion disk models \citep[e.g.,][]{Mummery_etal_2024_fited,Mummery_2025_TDEdisktheory}---the two methods have been shown to produce self-consistent black hole mass estimates to within $\sim0.5$ dex \citep{Angus_2026}. Therefore, we conservatively state that the TDE MBH has a mass of $\log(M_{\bullet}/M_{\odot}) \approx 6-7$. 

\begin{itemize}
\item The wandering MBH is substantially less massive than the SMBH expected in the nucleus of the host galaxy, for which we infer $\log(M_{\bullet}/M_{\odot}) = 8.38 \pm 0.42$ from the $M_{\bullet}$--$\sigma$ relation. This mass difference disfavors a gravitational-wave recoil origin \citep[e.g.,][]{Komossa_2012_recoilBHrev}. If the off-nuclear MBH were the remnant of a merger involving the central SMBH, its mass would be expected to be greater than that of the nuclear SMBH. The much lower MBH mass inferred from TDE~2025abcr therefore argues against this scenario. Also, the recoiling-SMBH scenario requires the absence of an SMBH in the galaxy nucleus. Although there is currently no definitive evidence for or against a central SMBH from radio or X-ray emission, or from AGN signatures, our KCWI data show a sharp rise in stellar velocity dispersion toward the nucleus, hinting that a central SMBH is most likely present. A dynamical detection of a nuclear SMBH would therefore rule out the recoil scenario. 

Moreover, for a \(10^{8.4}\,M_\odot\) nuclear SMBH, a solar-type main-sequence star has a nominal tidal radius smaller than the Schwarzschild radius, and observable TDEs from such stars are strongly suppressed by relativistic direct capture above \(M_\bullet \approx 10^{8}\,M_\odot\), except possibly for favorable combinations of black hole spin and orbital parameters \citep{Kesden2012,Law-Smith2017}. Although evolved stars can still be disrupted by more massive SMBHs, such events are expected to produce much slower, longer-lived transients than the otherwise standard UV/optical event observed here \citep{2012_Macleod}. We therefore conclude that a recoiling \(10^{8.4}\,M_\odot\) nuclear SMBH is unlikely to have produced this TDE.

\item Another possibility for the origin of the wandering MBH is dynamical ejection through three-body interactions in a nucleus hosting multiple MBHs, in which the lightest MBH is expelled from the center \citep[e.g.,][]{Volonteri_Perna_2005_ejectMBH, 2020_Naoz}. In such interactions, the escaping MBH is expected to leave with a speed comparable to the binary orbital velocity; representative values are of order a few \(10^2\) km~s\(^{-1}\), although extreme cases can reach several \(10^3\) km~s\(^{-1}\) for very tight binaries \citep[e.g.,][]{Hoffman_BH_dyn_kick2007, Hoffman_Loeb_2007, Gualandris_2008_MBHejectionreview, Bonetti_2018}.  
Based on the H Balmer lines, we do not find a significant velocity offset between the TDE emission lines and the systemic redshift of the host galaxy, measuring only \(7 \pm 27\) km~s\(^{-1}\). Although the IR lines show a blueshift of \(\sim 200\) km~s\(^{-1}\), this is more naturally explained by radiative-transfer effects \citep{Roth_Kasen_2018}. In the absence of detailed radiative-transfer modeling, however, we conservatively adopt an upper limit of \(<200\) km~s\(^{-1}\) on the line-of-sight velocity offset of the wandering MBH relative to the host galaxy. This constraint applies only to the line-of-sight component; the MBH could still have a larger transverse velocity, or it may have undergone substantial deceleration through dynamical friction and now be near apocenter. Thus, we cannot entirely rule out a dynamical-ejection origin.

\item A third possibility is that the wandering MBH was brought in by a galaxy merger. In this scenario, the satellite galaxy is tidally stripped, leaving behind a compact stellar core that remains bound to the MBH and can efficiently refill the loss cone, thereby sustaining a measurable TDE rate. We do not detect tidal tails or other obvious merger signatures in deep Legacy Survey DR10 imaging \citep{2019AJ....157..168D}, in our high-resolution \textit{JWST} image (Figure \ref{fig:jwstmiri}), or in an unsharp-masked version of the host-galaxy image constructed to enhance faint tidal structures (Figure \ref{fig:unsharp}). We therefore find no evidence for a recent major merger. 

A minor merger, however, remains a strong possibility \citep{Tremmel_2018_wandMBH}. Minor mergers can naturally produce wandering MBHs embedded within a compact dense cluster of stars without creating large-scale disturbances within the host galaxy \citep[e.g.,][]{Ricarte_2021_AGN_Wandering_BHs, Naab_2009_minormerg, Hopkins_2010_galmerg, Seth_Anil2014, 2021_Dodd,  patra2025jwstkeckobservationsoffnuclear}. Using the \citet{Raines_Volonteri_2015} black hole--galaxy stellar mass relation, a \(10^{6}-10^{7}\,M_\odot\) MBH would correspond to a progenitor satellite with stellar mass of order \(\sim 10^{10}\,M_\odot\). Relative to the \(\sim 10^{11.5}\,M_\odot\) host galaxy, this implies a mass ratio of \(\sim 1\!:\!30\), consistent with a minor merger. Subsequent tidal stripping could then plausibly leave behind only \(\lesssim 10^{8}\,M_\odot\) in stars bound to the wandering MBH, consistent with our inferred upper limit on the currently surviving stellar population. \citet{2026arXiv260210180S} reported possible disturbed host morphology in galaxy profile modeling.  Although the imaging analyses differ in their sensitivity to low-surface-brightness structures, that reported disturbance provides independent support for a merger-related history and is compatible with a minor-merger origin. A definitive detection of an underlying compact stellar cluster at the TDE position would strengthen this case. 
\end{itemize}

We note that our assessment of the origin scenarios above is broadly consistent with \citet{2026arXiv260210180S}: the mass contrast between the two black holes and massive host nucleus make gravitational-wave recoil unlikely, whereas both dynamical ejection and a minor-merger origin remain viable.

\section{Conclusion}
\label{sec:conclusion}

We have presented a multi-wavelength study of TDE~2025abcr, the second optically selected off-nuclear TDE. The TDE is located at a projected offset of $9.08 \pm 0.02$\,kpc from the host-galaxy center and shows broad H and He emission in both the optical and IR confirming a TDE-H-He classification. From \texttt{MOSFiT} modeling and luminosity-based scaling relations, we infer a TDE MBH mass of $\sim 10^{6}$--$10^{7}\,M_{\odot}$, far smaller than the $\sim 10^{8.4}\,M_{\odot}$ SMBH expected in the host nucleus.  
The TDE emission lines show asymmetric red wings, and in the case of IR lines, a blueshifted peak by $\sim 200~\rm km~s^{-1}$,  indicating line formation within a dense, electron-scattering-dominated reprocessing outflow. We also observe velocity evolution in the \ion{N}{3} + \ion{He}{2} emission complex, which we interpret as evidence for changing radiative transfer conditions in an evolving reprocessing layer. In addition, the IR SED shows a clear excess above a simple thermal continuum, with a slope shallower than the Rayleigh--Jeans expectation. The viable explanations are free--free emission from a dense, evolving reprocessing outflow, or an unresolved stellar population at the TDE location. 

These results support a picture in which TDE~2025abcr occurred around a wandering MBH most likely delivered by a minor galaxy merger, although dynamical ejection from the nucleus cannot yet be excluded. Future observations can directly test the origin of the IR excess: if it is powered by free--free emission, it should fade as the reprocessing layer weakens, whereas emission from a stellar cluster should remain constant. Late-time UV--optical--IR SED measurements will therefore be especially valuable. Additional spectroscopy, particularly of the Bowen fluorescence lines in the UV, could further clarify the evolution of the ionizing continuum and the physical state of the line-forming gas. 

TDE~2025abcr is only the second optically discovered off-nuclear TDE, after TDE~2024tvd. While TDE~2025abcr shows a clearly resolvable host-galaxy offset in ground-based data, TDE~2024tvd showed only an astrometric offset from the ground and required high-resolution \textit{HST} and \textit{JWST} imaging for confirmation. This suggests that TDEs with smaller offsets ($\lesssim 1''$), for which ground-based data do not provide robust offset measurements, are perhaps being classified as nuclear TDEs (e.g. TDE~2023mfm). A conservative lower limit on the optical off-nuclear TDE fraction is therefore $\sim 3/150 \approx 2\%$, although the true fraction could be higher. This estimate is compatible with \citet{2026arXiv260210180S}, who obtained a $90\%$-confidence lower limit of $0.2\%$ for the overall offset fraction and an upper limit of $9.6\%$ for highly offset events relative to the unresolved/nuclear TDE rate, while likewise emphasizing that subarcsecond offsets can be missed in ground-based observations. TDE~2024tvd and TDE~2025abcr highlight the potential of off-nuclear TDEs as probes of black hole demographics beyond galaxy centers and motivate systematic searches for similar events in the \textit{Roman}--LSST datasets.

\section*{Acknowledgments}

K.C.P. thanks Wenbin Lu for helpful discussions. The UCSC transients team is supported in part by STScI grant JWST-DD-12485 and by a fellowship from the David and Lucile Packard Foundation to R.J.F. E.R.L. and C.-P.M. acknowledge support of NSF AST-2206307, the Heising-Simons Foundation, and the Miller Institute for Basic Research in Science. J.L.W. acknowledges the support of NSF AST-2206219. E.R.R.\ acknowledges the Heising-Simons Foundation and NSF: AST--1852393, AST--2150255, and AST--2206243. 

This work is based in part on observations made with the NASA/ESA/CSA James Webb Space Telescope. The data were obtained from the Mikulski Archive for Space Telescopes (MAST) at the Space Telescope Science Institute, which is operated by the Association of Universities for Research in Astronomy, Inc., under NASA contract NAS 5--03127 for JWST. 

All \textit{JWST} data used in this paper can be found in MAST: \href{https://doi.org/10.17909/tas2-wf16}{doi:10.17909/tas2-wf16}. These observations are associated with program 12485(PI Patra). Support for program 12485 was provided by NASA through a grant from the Space Telescope Science Institute, which is operated by the Association of Universities for Research in Astronomy, Inc., under NASA contract NAS 5-03127.

Some of the data presented herein were obtained at Keck Observatory, which is a private 501(c)3 non-profit organization operated as a scientific partnership among the California Institute of Technology, the University of California, and the National Aeronautics and Space Administration. The Observatory was made possible by the generous financial support of the W.\ M.\ Keck Foundation.  The authors wish to recognize and acknowledge the very significant cultural role and reverence that the summit of Maunakea has always had within the Native Hawaiian community. We are most fortunate to have the opportunity to conduct observations from this mountain.

\bibliographystyle{mnras}

\bibliography{references_1, references_2, references_3, references_4, references_5}

\clearpage 

\begin{appendix}
\label{sec:appendix}

\section{\texttt{MOSFiT} Analysis}

\begin{figure}[!t]
    \centering
    \includegraphics[width=0.8\linewidth]{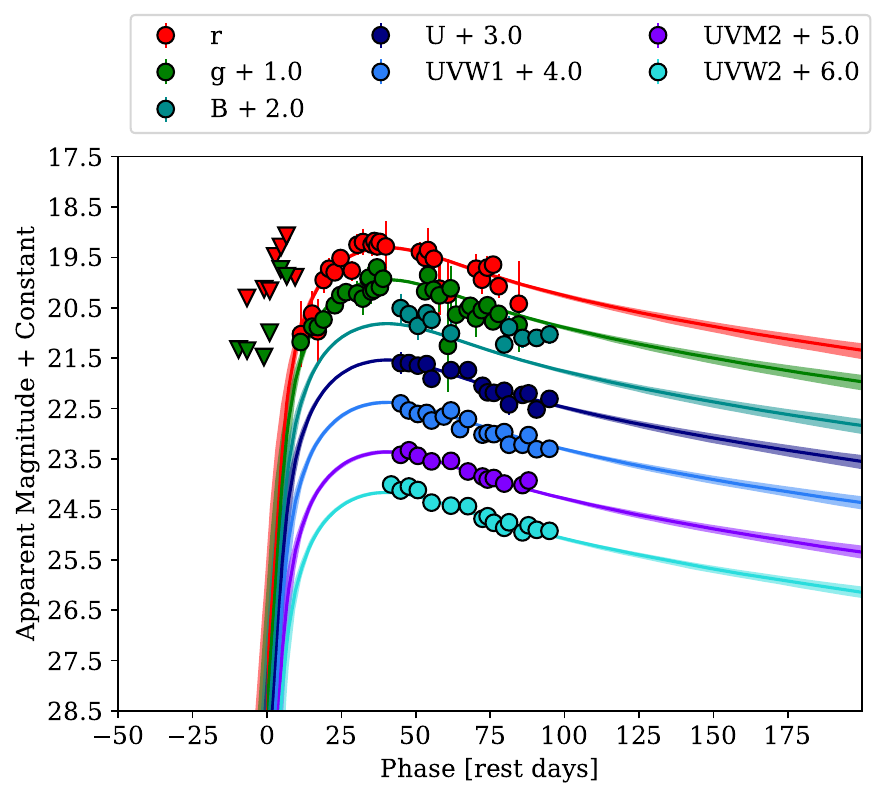}
    \caption{\texttt{MOSFiT} fitting to the UV and optical light curves of TDE~2025abcr. }
    \label{fig:mosfit}
\end{figure}

\begin{figure}[!t]
    \centering
    \includegraphics[width=\linewidth]{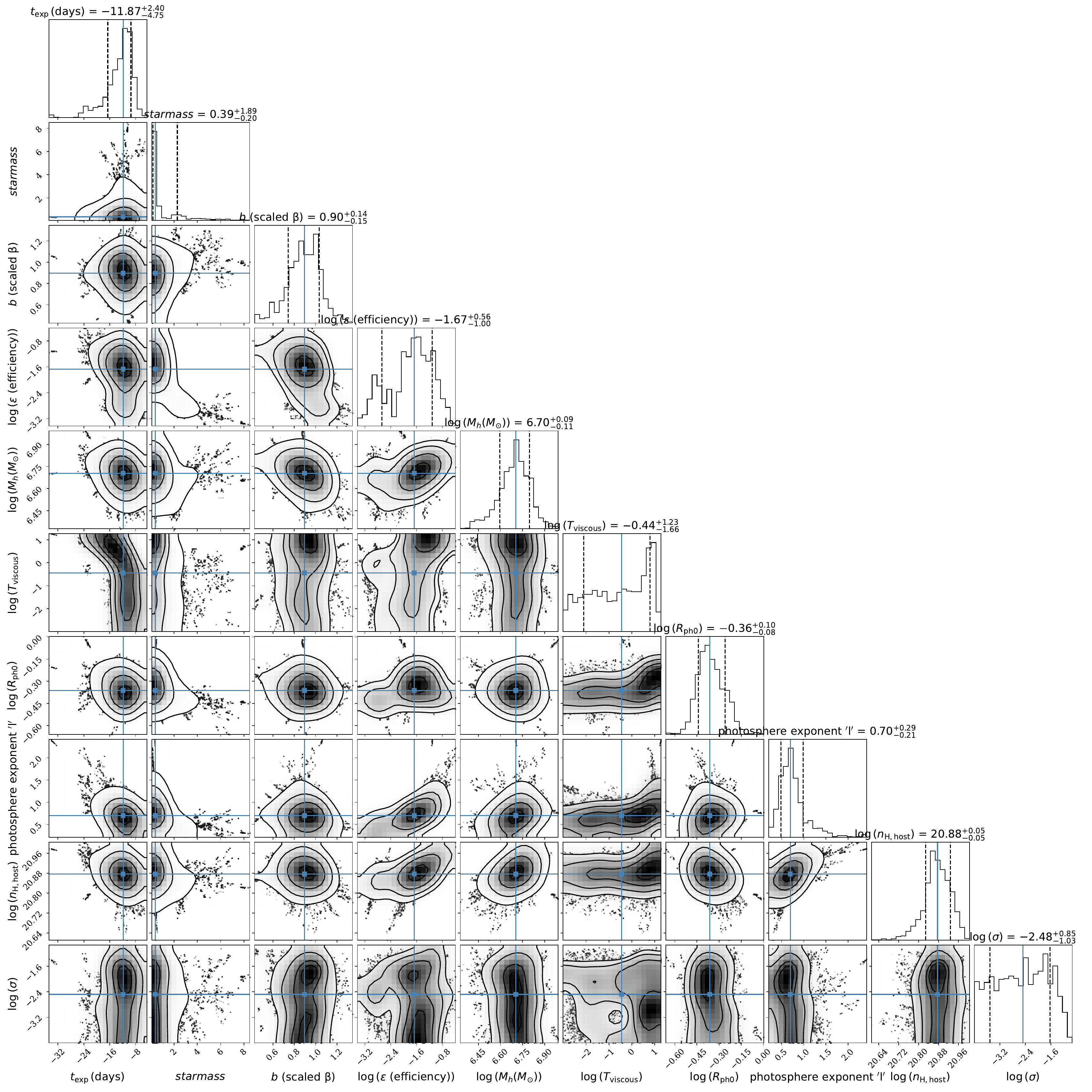}
    \caption{Posterior distributions of the fitted \texttt{MOSFiT} parameters. The solid blue line represents the median of the posterior distribution, while the dashed black lines show the 16th and 84th percentiles.}
    \label{fig:mosfitcorner}
\end{figure}

We modeled the \textit{Swift} UV and ZTF optical photometry with \texttt{MOSFiT}, which employs a Monte Carlo Markov Chain (MCMC) framework using the \texttt{emcee} sampler \citep{Foreman-Mackey_etal_2013_emcee}. We assessed convergence using the potential scale reduction factor, requiring a value $<1.2$ \citep{Gelman_Rubin_1992}; this criterion was satisfied after approximately 5000 steps with 150 walkers. The best-fit \texttt{TDE} model to the multi-band light curves and the posterior distributions of the fitted parameters are shown in Figures \ref{fig:mosfit} and \ref{fig:mosfitcorner}, respectively.

\section{Additional Figures}

\begin{figure*}[tbp]
    \centering
    \includegraphics[angle=-90, origin=c, width=0.6\linewidth,height=0.4\textheight]{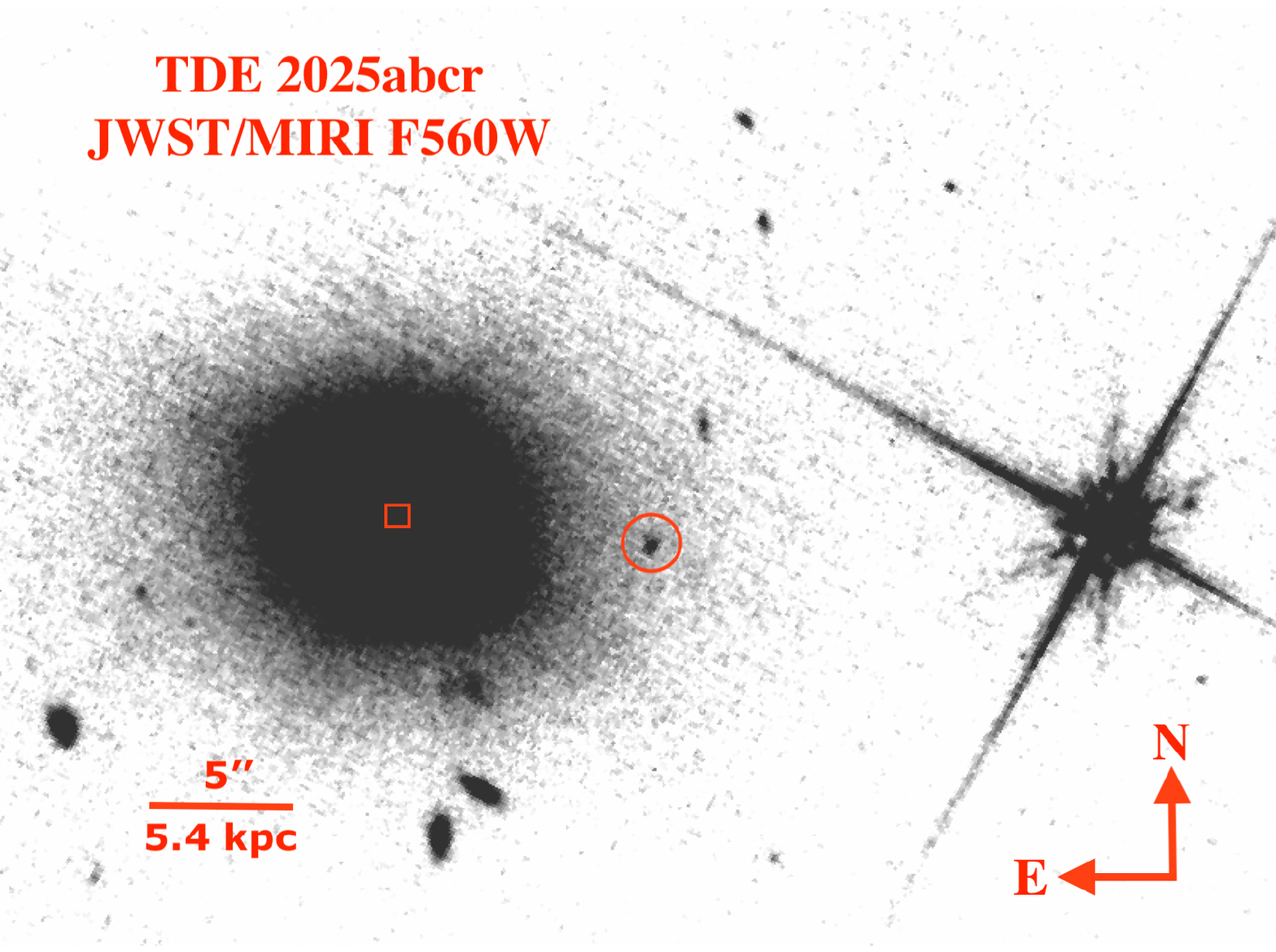}
    \caption {\textit{JWST} MIRI F560W image of TDE 2025abcr shows a rich galactic environment with several satellite galaxies. The TDE is marked with the red circle and the host galaxy’s nucleus
with the red square.}
    \label{fig:jwstmiri}
\end{figure*}

\begin{figure}[tbp]
    \centering
    \includegraphics[width=.8\linewidth]{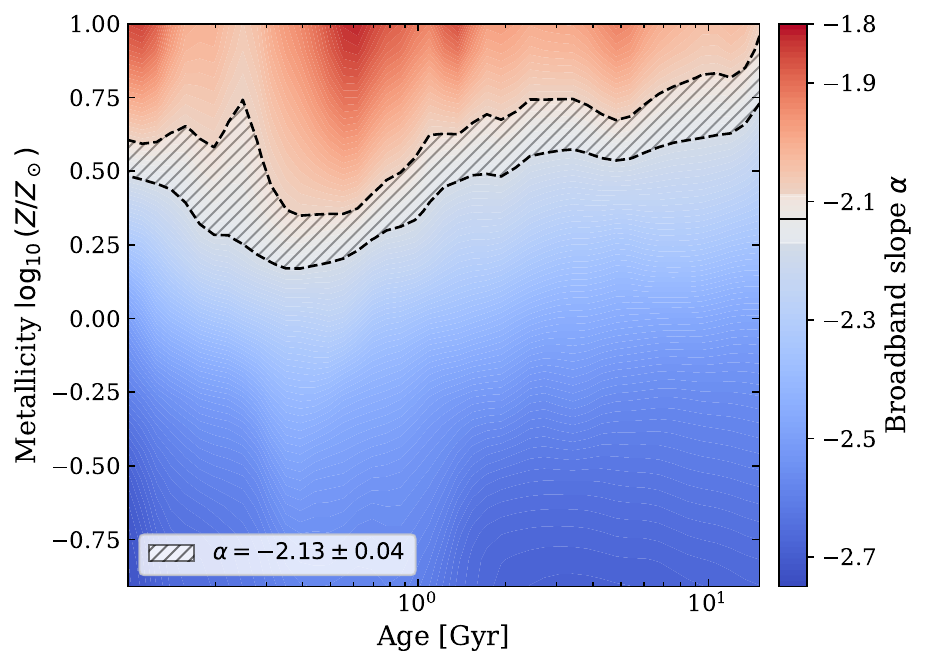}
    \caption{Broadband IR spectral slope of \texttt{FSPS} simple stellar populations as a function of age and metallicity, assuming a \citet{2001_Kroupa} IMF. The slope $\alpha$ is measured by fitting $\nu L_\nu\propto\lambda^\alpha$ over $2$--$7\,\mu{\rm m}$. The color background shows the model slope after mild smoothing for visualization; the hatched region denotes agreement with the observed value, $\alpha=-2.13\pm0.04$. To reproduce the observed IR slope in the \textit{JWST} data, the stellar cluster around the wandering MBH needs to have a supersolar metallicity. The vertical striations are likely excess emission contributions from short-lived, cool evolved stars, particularly red giants and asymptotic giant branch stars.}
    \label{fig:fsps_slope}
\end{figure}

\begin{figure}[tbp]
    \centering
    \includegraphics[width=\linewidth]{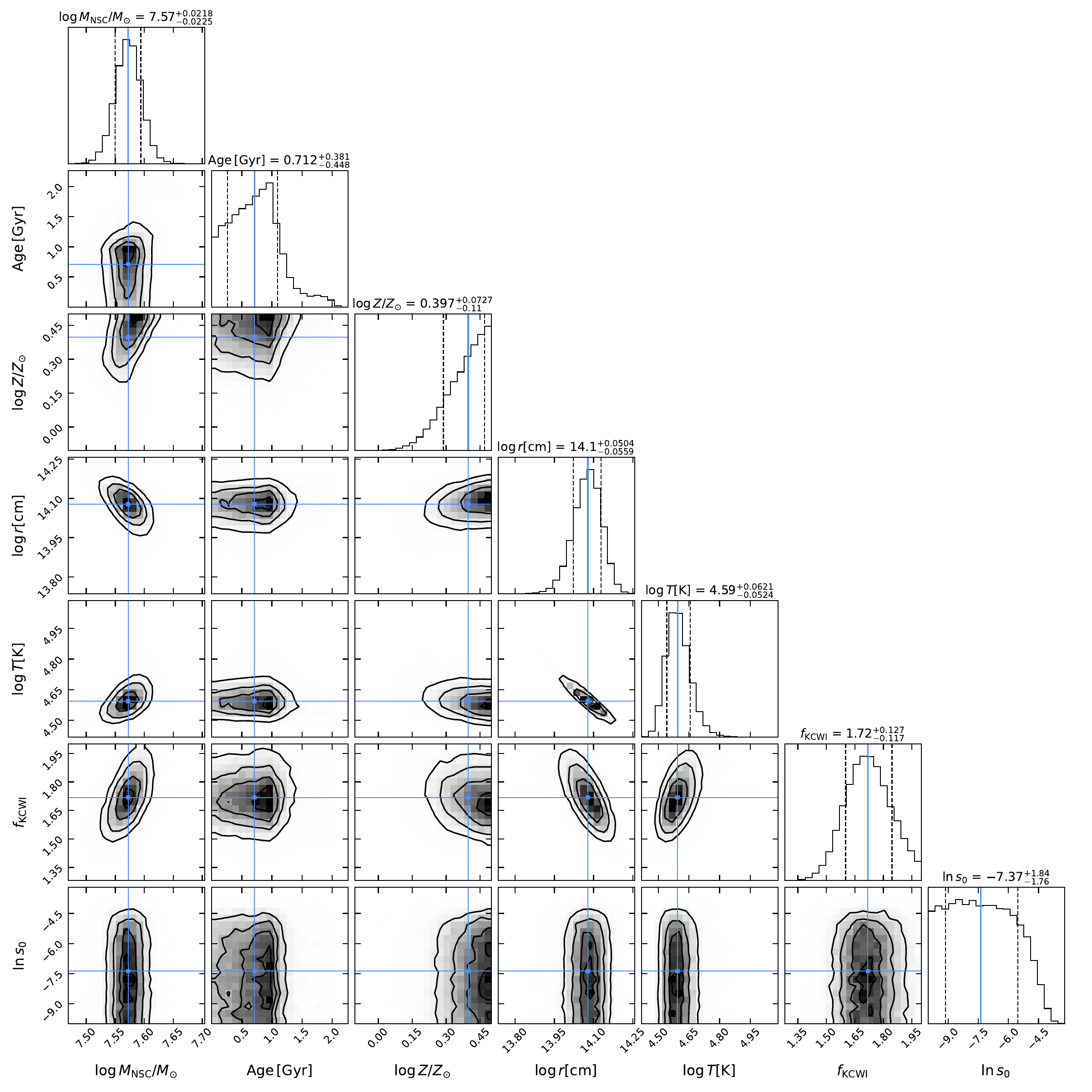}
    \caption{Posterior distributions of the fitted SED model parameters. The solid blue line represents the median of the posterior distribution, while the dashed black lines show the 16th and 84th percentiles.}
    \label{fig:sed_corner}
\end{figure}

\end{appendix}

\end{document}